\documentclass[dvips, useAMS, usenatbib]{mn2e}
\usepackage{amsbsy}
\usepackage{graphicx}
\usepackage{amssymb}
\usepackage{times}
\usepackage{color}

\makeatletter

\providecommand{\tabularnewline}{\\}

\newcommand{\Kelvin}{\mathrm{K}}
\newcommand{\Msun}{\mathrm{M_{\sun}}}

\newcommand{\Lsun}{\mathrm{L_{\sun}}}
\newcommand{\MsunPerYear}{\mathrm{M_{\sun}\,yr^{-1}}}

\newcommand{\aap}{A\&A}

\newcommand{\apj}{ApJ}
\newcommand{\apjl}{\apj}

\newcommand{\mnras}{MNRAS}
\newcommand{\apjs}{ApJS}

\newcommand{\nat}{Nature}
\newcommand{\apss}{Ap\&SS}
\newcommand{\araa}{ARA\&A}

\newbox\grsign \setbox\grsign=\hbox{$>$} \newdimen\grdimen \grdimen=\ht\grsign
\newbox\simlessbox \newbox\simgreatbox
\setbox\simgreatbox=\hbox{\raise.5ex\hbox{$>$}\llap
     {\lower.5ex\hbox{$\sim$}}}\ht1=\grdimen\dp1=0pt
\setbox\simlessbox=\hbox{\raise.5ex\hbox{$<$}\llap
     {\lower.5ex\hbox{$\sim$}}}\ht2=\grdimen\dp2=0pt

\def\simless{\mathrel{\copy\simlessbox}}

\makeatother

\begin{document}
\title[Radiation-driven outflow in AGN]{On the large-scale outflows in  
active galactic nuclei: consequences of coupling the mass-supply
  rate and accretion luminosity}

\author[R. Kurosawa and D. Proga.]{Ryuichi   Kurosawa\thanks{E-mail:rk@physics.unlv.edu},   and Daniel Proga\\ Department of Physics and Astronomy , University of   Nevada Las Vegas,  Box 454002,  4505 Maryland Pkwy,  Las Vegas, NV 89154-4002, USA}

\date{Dates to be inserted}


\maketitle

\begin{abstract}

We present two-dimensional hydrodynamical simulations of
\textcolor{black}{slowly rotating}  
gas that is under the influence of the gravity of a super massive black hole and
is irradiated by a thin UV accretion disc and a spherical X-ray corona. 
We calculate the accretion luminosity of a system based on
the accretion-rate which is assumed to be equal to the mass-supply rate at
the radius of $\sim10^{-2}$~pc. 
For the models with high temperature gas at large
radii ($\sim10$~pc) and high luminosities,
we find a strong correlation between the mass-outflow rate 
($\dot{M}_{\mathrm{out}}$) and the luminosity ($L$).
The power law index ($q$) describing the 
$\dot{M}_{\mathrm{out}}$--$L$ relation is $q=2.0\left(\pm0.1\right)$, 
which is very similar to that for radiation-driven stellar and 
disc wind models. More surprisingly, for high density
at large radii, we find steady state solutions with the accretion luminosity {\it exceeding} the
Eddington limit.
The super-Eddington accretion proceeds in the equatorial region and is possible 
because the radiation flux from the disc is significantly reduced in
the equatorial direction due to the geometrical foreshortening effect.
In all models, an outflow is driven from an inflow with sub-Keplerian rotation.  
For high temperature at large radii, the inflow occurs over a wide
range of the polar angles, and the outflow occurs in a relatively narrow polar cone.
However, for the super-Eddington cases with low temperature at large
radii, the inflow persists only very close to
the equatorial plane, resembling a thin accretion disc, while the outflow 
arises in a wide range of radii and polar angles. The geometry of this extreme 
inflow-outflow solution is very similar to a radiation-driven wind from a 
luminous Keplerian accretion disc. 
For the cold super-Eddington solutions,
$\dot{M}_{\mathrm{out}}$ is only very weakly
correlated with $L$, i.e. $0 \simless q \simless 0.2$.
This weaker correlation 
is mainly caused
by a mismatch between the direction of escaping photons and
the inflowing gas: 
the radiation is emitted mostly in the polar directions
whereas the inflowing gas occurs mainly in the equatorial region.

\end{abstract}

\begin{keywords} 

accretion, accretion discs -- hydrodynamics -- galaxies: kinematics
and dynamics -- methods: numerical -- galaxies: active 

\end{keywords}

\section{Introduction}

\label{sec:Introduction}

Induced by accretion of gas on to a super massive
($10^{6}$--$10^{10}\,\Msun$) black hole (SMBH), active galactic nuclei
(AGN) release a large amount of energy by radiating photons
($10^{10}$--$10^{14}\Lsun$). These photons influence the physical
properties (e.g.~the ionization structure, gas dynamics and density
distribution) of the vicinity of AGN, the AGN host galaxies, and the
inter-galactic media of galaxy clusters to which they belong
(e.g.~\citealt*{Quilis:2001}; \citealt{DallaVecchia:2004};
\citealt{McNamara:2005}; \citealt{Zanni:2005}; \citealt{Fabian:2006};
\citealt{Vernaleo:2006}).
Importance of the feedback from AGN has been recognised in many areas
of astrophysics, 
e.g.~(i)~co-evolution of galaxies and black holes
(e.g.~\citealt{ciotti:1997}, \citeyear{ciotti:2001},
\citeyear{ciotti:2007}; \citealt{king:2003}; \citealt{Hopkins:2005};
\citealt*{Murray:2005}; \citealt{Sazonov:2005};
\citealt*{DiMatteo:2005}; \citealt*{Springel:2005};
\citealt{Brighenti:2006}; \citealt{Fontanot:2006};
\citealt*{Wang:2006}, \citealt*{Tremonti:2007};
\citet*{Yuan:2009}; \citealt*{Ciotti:2009}), (ii)~star formation in host galaxies
(e.g.~\citealt{Hamann:1992}; \citealt{Maloney:1999};
\citealt*{Murray:2005}; \citealt*{Kim:2006};
\citealt{Schawinski:2009}; \citealt{Schawinski:2009b}), (iii)~the
so-called cooling flow problem  
in galaxy clusters (e.g.~\citealt{Binney:1995};
\citealt{Tucker:1997}; \citealt{Soker:2005};
\citealt*{Scannapieco:2005}; \citealt*{Rafferty:2006};
\citealt{McNamara:2007}; \citealt*{Guo:2008};
\citealt*{Rafferty:2008}).

The form of AGN feedback can be mechanical (e.g.~\citealt{Binney:1995};
\citealt{Begelman:2005}; \citealt*{Springel:2005b}) or radiative
in which photon momentum, energy or both are dynamically important
(e.g.~\citealt{ciotti:1997}, \citeyear{ciotti:2001}, \citeyear{ciotti:2007};
\citealt{Ciotti:2009}). The radiation can produce
mechanical feedback via radiation pressure on matter in the form of
outflows/winds (e.g.~\citealt{Proga:2007b}; \citealt*{Proga:2008};
\citealt*{Dorodnitsyn:2008a,Dorodnitsyn:2008b}; \citealt{Kurosawa:2008,kurosawa:2008b}).
A strong support for the existence of such (radiation-driven) winds
is often found in the highly blueshifted broad absorption line features
seen in the UV and optical spectra of AGN (e.g.~\citealt{Murray:1995b};
\citealt*{Proga:2000}; \citealt*{Chartas:2003}; \citealt*{Crenshaw:2003};
\citealt{Proga:2004}; \citealt{Hamann:2008}), provided the
ionization state of the gas is appropriate. We note that the outflows
can be also produced by magnetocentrifugal force (e.g.~\citealt{Blandford:1982};
\citealt*{Emmering:1992}; \citealt{Konigl:1994}; \citealt{Bottorff:1997}),
Poynting flux/magnetic towers (e.g.~\citealt*{Lovelace:1987}; \citealt{Lynden-Bell:1996,Lynden-Bell:2003};
\citealt{Li:2001}; \citealt{Proga:2003}; \citealt*{Kato:2004}; \citealt*{Nakamura:2006};
\citealt{Kato:2007}), and thermal pressure (e.g.~\citealt{Weymann:1982};
\citealt*{Begelman:1991}; \citealt{Everett:2007}). 

In our previous studies (\citealt{Proga:2007b}; \citealt{Proga:2008};
\citealt{kurosawa:2008b}, hereafter Papers~I, II and III, respectively),
we performed hydrodynamical simulations of radiatively driven AGN
outflows produced from the infalling gas. Our models
include radiation forces due to electron scattering and line processes,
and radiative cooling/heating. These models use a fixed accretion
luminosity during the whole simulations regardless of the mass inflow
rate through the model inner boundary. In principle,
the rate at which the gas is supplied to the region below the inner
boundary ($\sim10^{-2}$~pc) would influence the mass accretion rate
on to the SMBH at the centre and its luminosity. This could results
in models with an inconsistent amount of the radiative feedback for
a given amount of the mass supplied rate to the inner region of AGN.
Here we relax this assumption of
the constant luminosity, and instead dynamically couple the accretion
luminosity to the mass supply rate.
With this coupling, we examine the following two key issues in this study. 

Firstly, we investigate how the mass-outflow rate depends
on the self-consistently determined accretion luminosity (or equivalently
the Eddington ratio $\Gamma$ of the system). This provides a useful
information on the mass supply rate and the mass inflow morphology
which can be incorporated with a smaller scale disc accretion model
(e.g.~\citealt{Ohsuga:2007}). Similarly, the mass-outflow rates
predicted here can be incorporated with a larger scale galaxy
evolution model (e.g.~\citealt{Ciotti:2009}).

Secondly, we test whether
steady state solutions with a super-Eddington luminosity  are possible when
the main radiation source has a disc geometry instead of a spherical
geometry. The system with the disc radiation field has a possible channel
of a high mass supply rate along a equatorial plane at
a large distance from the disc itself. Because of the disc geometry,
the radiation would peak strongly along the pole or disc axis directions,
and it would gradually decrease toward equatorial direction hence the
radiation force is much weaker in the equatorial plane.  

In the following section, we describe our method and model assumptions.
We give the results of our hydrodynamical simulations in Section~\ref{sec:Results}.
Discussions on accretion flows with a super-Eddington luminosity and
the effect of artificially limiting luminosity are given in Section~\ref{sec:Discussions}.
Finally, the conclusions of this study are summarised in Section~\ref{sec:Conclusions}.

\section{Method}

\label{sec:Method}

\subsection{Overview}

\label{sub:Overview}

We mainly follow the methods used for the axisymmetric models by \citet{Proga:2000}
and \citet{Proga:2004}, but with a change in the treatment of the
accretion luminosity (Section~\ref{sub:Coupling-of-Mass-Accretion}).
The model geometry and the assumptions of the SMBH and the disc are
very similar to those in Papers~I, II and III. 
Readers are referred to Fig\@.~1 of Paper~III
for the diagram of our basic model configuration. A SMBH with its
mass $M_{\mathrm{BH}}$ and its Schwarzschild radius $R_{\mathrm{S}}=2GM_{\mathrm{BH}}/c^{2}$
is placed at the centre of the polar coordinate system ($r$, $\theta$).
The X-ray emitting corona regions is defined as a sphere with its
radius $r_{*}$ which surrounds the SMBH. The geometrically thin and
optically thick accretion disc (e.g.~\citealt{shakura:1973}) is
placed on the equatorial plane ($\theta=\pi/2$ plane). The normal
vector of the disc coincides with the symmetry axis. The hydrodynamic
simulations will be performed in the polar coordinate system (assuming
axisymmetry) with $r$ between the inner boundary $r_{\mathrm{i}}$
and the outer boundary $r_{\mathrm{o}}$. The radiation forces, from
the corona region (the sphere with its radius $r_{*}$) and the accretion
disc, acting on the gas are assumed to be radial only. The point-source
like approximation for the disc radiation pressure is used here since
the accretion disc radius ($r_{\mathrm{D}}$) is assumed to be much
smaller than the inner radius, i.e.~$r_{\mathrm{D}}\ll r_{\mathrm{i}}$.

Because of the assumed flat accretion disc geometry, the disc radiation
flux (${\cal F}_{{\rm \mathrm{disc}}}$) peaks in the direction of
the disc rotational axis, and it gradually decreases as the polar
angle $\theta$ increases, i.e.~${\cal F}_{{\rm \mathrm{disc}}}\propto|\cos{\theta}|$.
The flow is also irradiated by the corona which is assumed to be spherical.
Further, we assume that the total accretion luminosity $L$ consists
of two components: (1)~$L_{{\rm disc}}=f_{{\rm disc}}L$ due to the
accretion disc and (2)~$L_{\ast}=f_{{\rm \ast}}L$ due to the corona 
\textcolor{black}{(i.e.~$f_{\mathrm{disc}} + f_{\ast} = 1$).}
We assume that the disc emits only UV photons, whereas the corona
emits only X-rays, i.e.~the system UV luminosity, $L_{{\rm UV}}=f_{{\rm UV}}L=L_{{\rm disc}}$
and the system X-ray luminosity, $L_{{\rm X}}=f_{{\rm X}}L=L_{\ast}$
(in other words $f_{{\rm UV}}=f_{{\rm disc}}$ and $f_{{\rm X}}=f_{\ast}$).


\textcolor{black}{
  We include the effect of radiation forces due to
  electron scattering and line processes (line
  force).  To evaluate the line force, we follow the method described
  by \citet{Proga:2000} who use a modified version of the formalism
  developed by \citet*{Castor:1975} (CAK model).
  In our model, the line force is evaluated by using the analytical
  formulae of \citet{Stevens:1990} who parametrised the force in
  terms of the photoionization parameter, defined as $\xi=4\pi
  {\cal F}_{\mathrm{X}}/n$ where  ${\cal F}_{\mathrm{X}}$ and $n$ are
  the local X-ray flux and the number density of the gas,
  respectively.  This parametrisation of the line force is
  computationally efficient and it provides good estimates for
  the number and opacity distribution of spectral lines for a given
  $\xi$.}

\textcolor{black}{
  Note that we account for the attenuation of X-ray
  radiation by computing X-ray optical depth in radial direction;
  however, we assume that the gas is optically thin in UV. The
  assumption of no attenuation of UV radiation has to be made in our
  formalism since the photoionization model of \citet{Stevens:1990} does
  not include the UV attenuation; therefore, to be consistent with the
  their model. We will discuss the validity of this
  assumption later in Section~\ref{sub:thin_UV}. 
}

\textcolor{black}{
  With the simplification above, only the corona
  (X-ray) radiation contributes to bringing atoms to very high
  ionization states. We assume that the corona contributes to the
  radiation force due to electron scattering, but not to the line
  force although some metal lines in the soft X-ray may contribute to
  the line force in some cases. On the other hand, the disc radiation
  (UV) contributes significantly to the line force and the force
  due to electron scattering.
}

\textcolor{black}{In all the models presented in this work, we assume that the inflowing gas
  at the outer boundary is very slowly rotating, i.e. the rotation
  velocity of the gas is much smaller than the Keplerian velocity. We
  use the same specific angular momentum distribution of the gas at
  the outer boundary as in Paper III.  
}

\subsection{Hydrodynamics}

\label{sub:Hydrodynamics}

We employ hydrodynamical (HD) simulations of the outflow from accretion
on to a central part of AGN in 2-D (axisymmetric), using the {\sc ZEUS-MP}
code \citep[cf.][]{Hayes:2006} which is a massively parallel MPI-implemented
version of the {\sc ZEUS-3D} code (cf., \citealt{Hardee:1992}; \citealt{Clarke:1996}).
The {\sc ZEUS-MP} is a Eulerian hydrodynamics code which uses the method
of finite differencing on a staggered mesh with a second-order-accurate,
monotonic advection scheme \citep{Hayes:2006}. To compute the structure
and evolution of a flow irradiated by a strong continuum radiation
of AGN, we solve the following set of HD equations: \begin{eqnarray}
\frac{D\rho}{Dt}+\rho\,\boldsymbol{\nabla}\cdot\boldsymbol{v} & = & 0,\label{eq:hydro01}\end{eqnarray}

\begin{equation}
\rho\frac{D\boldsymbol{v}}{Dt}=-\boldsymbol{\nabla}P+\rho\,\boldsymbol{g}+\rho\,\boldsymbol{g}_{\mathrm{rad}},\label{eq:hydro02}\end{equation}

\begin{equation}
\rho\frac{D}{Dt}\left(\frac{e}{\rho}\right)=-P\,\boldsymbol{\nabla}\cdot\boldsymbol{v}+\rho\,\mathcal{C},\label{eq:hydro03}\end{equation}
 where $\rho$, $e$, $P$ and $\boldsymbol{v}$ are the mass density,
energy density, pressure, and the velocity of gas respectively. Also,
$\boldsymbol{g}$ is the gravitational force per unit mass. The Lagrangian/co-moving
derivative is defined as $D/Dt\equiv\partial/\partial t+\boldsymbol{v}\cdot\boldsymbol{\nabla}$.
We have introduced two new components to the {\sc ZEUS-MP} in order to treat
the gas dynamics more appropriate for the flow in and around AGN.
The first is the acceleration due to radiative force per unit mass
($\boldsymbol{g}_{\mathrm{rad}}$) in equation~\ref{eq:hydro02},
and the second is the the effect of radiative cooling and heating, 
simply as the net cooling rate ($\mathcal{C}$) in equation~\ref{eq:hydro03}.
We assume the equation of state to be in the form of $P=\left(\gamma-1\right)e$
where $\gamma$ is the adiabatic index, and $\gamma=5/3$ for all
the models presented in this paper. Our numerical method used in this
paper are identical, in most aspects, to that described in Papers~I
and II. 
Next, we briefly discuss our implementation of $\boldsymbol{g}_{\mathrm{rad}}$
and $\mathcal{C}$. 


\subsection{Radiation force and radiative heating/cooling}
\label{sub:rad_force_and_heating}

\textcolor{black}{
  As mentioned earlier, we consider two different continuum radiation
  sources in our models: (1)~the accretion disc, and (2)~the central
  spherical corona. 
  In the point-source approximation limit, the radiation flux from the
  central X-ray corona region can be written as 
  \begin{equation}
    \mathcal{F}_{*}=\frac{L_{*}}{4\pi r^{2}}
    \label{eq:corona-flux}
  \end{equation}
  where $r$ is the radial distance from the centre (by neglecting
  the source size). Here we neglect the geometrical obscuration of the
  corona emission by the accretion disc and vice versa. On the other
  hand, the disc radiation depends on the polar angle $\theta$ because
  of the source geometry. Again following Papers~I and II (see also
  \citealt{proga:1998}), the disc flux $\mathcal{F}_{\mathrm{disc}}$ is assumed
  to be radial and $\mathcal{F}_{\mathrm{disc}}\propto\left|\cos\theta \right|$. 
  Consequently, the disc radiation flux at a distance $r$ from
  the centre can be written as 
  \begin{equation}
    \mathcal{F}_{\mathrm{disc}}=2\,\left|\cos\theta\right|\,\frac{L_{\mathrm{disc}}}{4\pi
      r^{2}}
    \label{eq:disk-flux}
  \end{equation}
  where $\theta$ is the polar angle (angle between the disc normal and the position
  vector $\boldsymbol{r}$). The leading term $2$ in this expression
  comes from the normalisation of the flux. 
}

\textcolor{black}{
  To evaluate the radiative acceleration due to line force,
  we follow the method in \citet{Proga:2000} (see also
  \citealt{proga:1998}) who applied the modified 
  version of CAK model (see also \citealt{Castor:1975}).
  Their model works under the assumption of the Sobolev approximation
  (e.g., \citealt{Sobolev:1957}; \citealt{Castor:1970};
  \citealt{Lucy:1971}). Following \citet{Proga:2000}, the radiative acceleration of a unit
  mass at a point $\boldsymbol{r}$ can be written as 
  \begin{equation}
    \boldsymbol{g}_{\mathrm{rad}}=\oint_{\Omega}\left[1+\mathcal{M}\right]\left[\frac{\sigma_{e}I\left(\boldsymbol{r},\hat{\boldsymbol{n}}\right)}{c}\right]\hat{\boldsymbol{n}}\,
    d\Omega 
    \label{eq:rad_force}
  \end{equation}
  where $I$, $\Omega$ and $\sigma_{e}$ are the frequency-integrated
  continuum intensity in the direction $\hat{\boldsymbol{n}}$, the
  solid angle subtended by the source of continuum radiation, and the
  electron scattering cross section, respectively.
  The force multiplier $\mathcal{M}$
  is a function of optical depth parameter $\tau'$ which is similar
  to the Sobolev optical depth (c.f.~\citealt{Rybicki:1978}). 
  The parameters in $\mathcal{M}$ are
  functions of the photoionization parameter $\xi$
  (\citealt{Stevens:1990}). 
  Using eqs.~(\ref{eq:corona-flux}),
  (\ref{eq:disk-flux}), (\ref{eq:rad_force}), and the luminosity ratios ($f_{*}$ and
  $f_{\mathrm{disc}}$), the radiative acceleration term in
  equation~\ref{eq:hydro02} can be reduced to 
  \begin{equation}
    \boldsymbol{g}_{\mathrm{rad}}=\frac{\sigma_{e}L}{4\pi r^{2}c}\left[
    f_{*}+2\,f_{\mathrm{disc}} \left( 1 + \mathcal{M} \right) \left|\cos\theta\right|\, \right]
    \,\boldsymbol{\hat{r}}\,.
    \label{eq:rad-force-final}
  \end{equation}
}

\textcolor{black}{
To evaluate the gas temperature, we follow the method described by
\citet{Proga:2000} and \citet{Proga:2004}. The model includes some effects of
photoionization. The gas is assumed to be optically thin to its own
cooling radiation. The radiative processes included are Compton
heating/cooling, X-ray photoionization and recombination,
bremsstrahlung, and line cooling. The net cooling rate depends on the
photoionization parameter $\xi$ and the characteristic temperature of
the X-ray radiation ($T_{\mathrm{X}}$). For more detail, readers are
referred to Papers~I and II (see also \citealt{Proga:2000}) for our
implementations of $\boldsymbol{g}_{\mathrm{rad}}$ 
and $\mathcal{C}$.
}

\textcolor{black}{
  Note that the angular distribution of the radiation in
  equation~\ref{eq:rad-force-final} and also the luminosity ratios
  ($f_{*}$ and $f_{\mathrm{disc}}$) are fixed during the whole
  simulations.  In other words,  the radiation from the disc
  (equation~\ref{eq:disk-flux}) and the X-ray corona region
  (equation~\ref{eq:corona-flux}) are treated as inner boundary 
  conditions; hence, they are not self-consistently determined while
  the simulations proceed. 
}

\subsection{Coupling of Mass-Accretion Rate and Luminosity }

\label{sub:Coupling-of-Mass-Accretion}

In our previous models (Papers~I, II and III), the accretion luminosity
of the system is kept fixed. Consequently, the radiative feedback
may not be consistent with the mass supplied to the inner region of
AGN transferred from our inner boundary. To overcome this possible
inconsistency of the accretion luminosity and the mass inflow rate,
we relax the assumption of the constant luminosity. The luminosity
is now coupled to the mass-accretion rate ($\dot{M}_{\mathrm{a}}$)
in the following way. At each time step, the accretion luminosity
is adjusted based on the history of the mass accretion rate through
the inner boundary. However, since our inner boundary is rather large
compare to $r_{*}$, we introduce a lag time ($\tau$). The lag time
roughly corresponds to the accretion time scale for the gas to reach
the BH from the inner boundary of our computational domain. This should
naturally cause a self-regulation of mass-accretion rate and the amount
of radiative feedback, i.e.~a higher mass accretion rate hence a
higher accretion luminosity will more strongly push surrounding
gas outward, and will naturally slow down the accretion process.

The mass accretion rate of the BH with mass $M_{\mathrm{BH}}$ at
a given time $t$ is $\dot{M}_{\mathrm{a}}\left(t\right)$, the accretion
luminosity is given by 

\begin{equation}
L\left(t\right)=\frac{2\eta\, G\, M_{\mathrm{BH}}\dot{M}_{\mathrm{a}}\left(t\right)}{R_{\mathrm{S}}}\label{eq:Lacc_time_dependet}\end{equation}
 where $\eta$, $G$ and $R_{\mathrm{S}}$ are the rest mass conversion
efficiency, the gravitational constant, and the Schwarzschild radius
of the BH, respectively. The growth rate of black hole mass is negligible
here in the time scale used in our simulation. We estimate the mass
accretion rate at a given time $t$ by computing the time average
of the mass inflow rate ($\dot{M}$) at the inner boundary
as our simulation proceeds. If the time duration used for the averaging
the mass accretion rate is $\Delta t$, then we can write the average
mass accretion rate ($\dot{M}_{\mathrm{a}}$) as 

\begin{equation}
\dot{M}_{\mathrm{a}}\left(t\right)=\frac{\int_{t-\tau-\Delta t}^{t-\tau}\dot{M}\left(t'\right)dt'}{\int_{t-\tau-\Delta t}^{t-\tau}dt'}\,.\label{eq:Mdot_acc_time_average}\end{equation}
In our models we set $\Delta t=\tau$, and the denominator simply
becomes $\tau$. In the `standard' $\alpha$ disc models, the accretion
timescale ($t_{\mathrm{acc}}$) or the lag time $\tau$ can be approximated
as $t_{\mathrm{acc}}=\mathcal{O}\left(t_{\mathrm{dyn}}/\alpha\right)$
where $t_{\mathrm{dyn}}$ and $\alpha$ are the dynamical timescale
at the outer radius ($r_{\mathrm{d}}$) of the disc and the $\alpha$
viscosity parameter (\citealt{shakura:1973}).
In this paper, we assume $r_{d}=2.6\times10^{16}$cm and $\alpha=0.1$,
which give $t_{\mathrm{dyn}}\approx10^{8}$s, and consequently
$\tau=t_{\mathrm{acc}}\approx10^{9}$~s (see also \citealt{ciotti:2001}
for a similar treatment). Note that the time scale for the
free-falling gas ($t_{\mathrm{ff}}$) 
to reach the centre from the inner boundary of our computational domain
is approximately $10^{8}$~s which is same order of magnitude as
the dynamical timescale mentioned above. For the models which has
a steady state solutions, we find that the basic flow properties
(e.g.~morphology, mass-accretion rate, and gas temperature) are rather
insensitive to the value of $\tau$. 

A caveat in the method described here is that we assume all the inflowing
gas which crosses the inner boundary reaches the accretion disc and
contributes to the accretion luminosity.  In reality, some
fraction of the gas which falls inward of the inner boundary
may not reach the accretion disc due to, again, strong disc radiation
(e.g.~\citealt{Proga:2000}).

\subsection{Model Setup}

\label{sub:Model-Setup}

The following parameters are common to all the models presented here,
and are exactly same as in Papers~I, II and III. We assume that
the central BH is non-rotating and has mass $M_{\mathrm{BH}}=10^{8}\, M_{\odot}$.
The size of the disc inner radius is assumed to be
$r_{*}=3\,R_{\mathrm{S}}=8.8\times10^{13}\,\mathrm{cm}$. The rest mass
conversion efficiency ($\eta$) in equation~\ref{eq:Lacc_time_dependet} is
assumed to be $0.0833$. The Eddington luminosity of the Schwarzschild
BH, i.e.~$4\pi cGM_{\mathrm{BH}}/\sigma_{e}$ for our system is about
$1.3\times10^{46}\,\mathrm{erg\, s^{-1}}$ or $3.3\times10^{8}\, \Lsun$.
The fractions of the luminosity in the UV ($f_{\mathrm{UV}}$) and
that in the X-ray ($f_{\mathrm{X}}$) are fixed at $0.95$ and $0.05$
respectively, as in Paper~I (their Run~C) and in Paper~II (their
Run~Cr). These two parameters determine the shape of the underlying
continuum emission of the central source (cf.~Section~\ref{sub:Hydrodynamics}).

The following ranges of the coordinates are adopted: $r_{\mathrm{i}}\leq r\leq r_{\mathrm{o}}$,
$0\leq\theta\leq\pi/2$ where $r_{\mathrm{i}}=500\, r_{*}$ and $r_{\mathrm{o}}=2.5\times10^{5}\, r_{*}$.
The radius of the central and spherical X-ray corona region $r_{*}$
coincides the inner radius of the the accretion disc. In our simulations,
the polar angle range is divided into 64 zones, and are equally spaced.
In the $r$ direction, the gird is divided into 140 zones in which
the zone size ratio is fixed at $\Delta r_{k+1}/\Delta r_{k}=1.04$.

For the initial conditions, the density and the temperature of gas
are set uniformly, i.e.~$\rho=\rho_{o}$ and $T=T_{o}$ everywhere
in the computational domain (Table~\ref{tab:Model-Summary}). The
initial velocity of the gas is assigned by assuming the same specific
angular momentum distribution used in Papers II and III. At the inner
and outer boundaries ($r_{\mathrm{i}}$ and $r_{\mathrm{o}}$), we apply
the outflow (free-to-outflow) boundary 
conditions, in which the field values are extrapolated beyond the
boundaries using the values of \emph{the ghost zones} residing outside
of normal computational zones (see \citealt{Stone:1992} for more
details). At $r=r_{\mathrm{o}}$, all HD quantities are assigned to the
initial conditions (e.g.~$T=T_{o}$ and $\rho=\rho_{o}$) during  the evolution of
each model; however, this outer boundary condition is applied only
when the gas is inflowing at $r=r_{\mathrm{o}}$, i.e.~when $v_{r}<0$.
The radial component of the velocity is allowed to float (unconstrained)
when $v_{r}>0$ at $r=r_{\mathrm{o}}$. In Paper~II, we also applied these
conditions to represent a steady flow condition at $r=r_{\mathrm{o}}$.
We found that this technique leads to a solution that relaxes to
a steady state in both spherical and non-spherical accretion with
an outflow (see also \citealt{Proga:2003b}). This imitates the condition
in which a continuous supply of gas is available at
$r=r_{\mathrm{o}}$. We also consider a several cases in which the
temperature at $r=r_{\mathrm{o}}$ is unconstrained (Section~\ref{sub:To_not_fixed}).

\subsection{Reference Values}

\label{sub:Reference-Values}

Important reference physical quantities relevant to our systems are
as follows. The Compton radius, $R_{\mathrm{C}}\equiv GM_{\mathrm{BH}}\mu\, m_{p}/kT_{\mathrm{C}}$,
is $8\times10^{18}\,\mathrm{cm}$ or equivalently $9\times10^{4}\, r_{*}$
where $T_{\mathrm{C}}$, $\mu$ and $m_{p}$ are the Compton temperature, the
mean molecular weight of gas and the proton mass, respectively. We
assume $T_{\mathrm{C}}=2\times10^{7}\,\Kelvin$ and $\mu=1$. The corresponding
speed of sound at infinity is $c_{\infty}=(\gamma kT_{\mathrm{C}}/\mu m_{p})^{1/2}=4\times10^{7}\,\mathrm{cm\, s^{-1}}$.
The corresponding Bondi radius \citep{Bondi:1952} is $R_{\mathrm{B}}=GM_{\mathrm{BH}}/c_{\infty}^{2}=4.8\times10^{18}\,\mathrm{cm}$
while its relation to the Compton radius is $R_{\mathrm{B}}=\gamma^{-1}R_{\mathrm{C}}$.
The Bondi accretion rate (for the isothermal flow) is $\dot{M}_{\mathrm{B}}=3.3\times10^{25}\,\mathrm{g\, s^{-1}}=0.52\,\MsunPerYear$.
The corresponding free-fall time ($t_{\mathrm{ff}}$) of gas from
the Bondi radius to the inner boundary is $2.1\times10^{11}\,\mathrm{sec}=7.0\times10^{3}\,\mathrm{yr}$.
The escape velocity from the inner most radius ($r_{\mathrm{i}}=500\, r_{*}$)
of the computational domain is about $7.7\times10^{4}\,\,\mathrm{km\, s^{-1}}$.

\begin{table*}

\caption{Summary of models with different combinations of $\rho_{o}$
and $T_{\mathrm{o}}$}

\label{tab:Model-Summary}

\begin{tabular}{rccccccc}
\hline 
 & $\rho_{\mathrm{o}}$ & $T_{\mathrm{o}}$ & $\dot{M}_{\mathrm{in}}\left(r_{\mathrm{i}}\right)$ & $\Gamma$ & $\dot{M}_{\mathrm{out}}\left(r_{\mathrm{o}}\right)$ & Variability & Outflow Morphology\tabularnewline
Model & $\left(10^{-21}\,\mathrm{g\, cm^{-3}}\right)$ & $\left(10^{7}\,\mathrm{K}\right)$ & $\left(10^{25}\mathrm{\, g\, s^{-1}}\right)$ & $\cdots$ & $\left(10^{25}\,\mathrm{g\, s^{-1}}\right)$ & $\cdots$ & $\cdots$\tabularnewline
\hline 
1 & 1 & 0.2 & 5.1$\left(0.002\right)^{\ddagger}$ & 0.30$\left(0.001\right)^{\ddagger}$ & 0.0$\left(0.0\right)^{\ddagger}$ & steady & no outflow\tabularnewline
2 & 2 & 0.2 & 8.6$\left(0.14\right)$ & 0.51$\left(0.01\right)$ & 1.4$\left(1.3\right)$ & steady & polar \tabularnewline
3 & 3 & 0.2 & 11$\left(0.36\right)$ & 0.65$\left(0.02\right)$ & 3.4$\left(2.1\right)$ & semi-steady & wide polar\tabularnewline
4 & 5 & 0.2 & 15$\left(1.8\right)$ & 0.89$\left(0.11\right)$ & 6.2$\left(3.1\right)$ & semi-steady & wide polar\tabularnewline
5 & 10 & 0.2 & 22$\left(2.3\right)$ & 1.3$\left(0.1\right)$ & 13$\left(4.6\right)$ & semi-steady & wide polar\tabularnewline
6 & 20 & 0.2 & 36$\left(3.8\right)$ & 2.1$\left(0.2\right)$ & 17$\left(7.7\right)$ & semi-steady & disc wind\tabularnewline
7 & 50 & 0.2 & 63$\left(8.6\right)$ & 3.8$\left(0.5\right)$ & 21$\left(16\right)$ & semi-steady & disc wind\tabularnewline
8 & 100 & 0.2 & 93$\left(6.2\right)$ & 5.5$\left(0.4\right)$ & 18$\left(13\right)$ & semi-steady & disc wind\tabularnewline
9 & 200 & 0.2 & 130$\left(5.0\right)$ & 7.7$\left(0.3\right)$ & 20$\left(12\right)$ & semi-steady & disc wind\tabularnewline
$\cdots$ & $\cdots$ & $\cdots$ & $\cdots$ & $\cdots$ & $\cdots$ & $\cdots$ & $\cdots$\tabularnewline
10 & 0.4 & 2 & 2.4$\left(0.01\right)$ & 0.14$\left(0.01\right)$ & 0.0$\left(0.0\right)$ & steady & no outflow\tabularnewline
11 & 1 & 2 & 8.3$\left(0.36\right)$ & 0.49$\left(0.02\right)$ & 2.0$\left(2.3\right)$ & semi-steady & narrow conic\tabularnewline
12 & 2 & 2 & 15$\left(6.5\right)$ & 0.89$\left(0.38\right)$ & 9.4$\left(11\right)$ & semi-steady & conic\tabularnewline
13 & 4 & 2 & 26$\left(13\right)$ & 1.5$\left(0.8\right)$ & 24$\left(22\right)$ & semi-steady & wide conic\tabularnewline
14 & 10 & 2 & 56$\left(31\right)$ & 3.3$\left(1.8\right)$ & 66$\left(91\right)$ & semi-steady & wide conic\tabularnewline
$\cdots$ & $\cdots$ & $\cdots$ & $\cdots$ & $\cdots$ & $\cdots$ & $\cdots$ & $\cdots$\tabularnewline
15 & 0.5 & 20 & 1.3$\left(4.9\right)$ & 0.074$\left(0.292\right)$ & 3.4$\left(4.5\right)$ & spiky & almost spherical\tabularnewline
16 & 1 & 20 & 2.8$\left(5.6\right)$ & 0.17$\left(0.33\right)$ & 4.3$\left(4.2\right)$ & spiky & spherical/polar\tabularnewline
17 & 1.5 & 20 & 7.5$\left(13\right)$ & 0.45$\left(0.77\right)$ & 7.1$\left(9.1\right)$ & spiky & polar\tabularnewline
18 & 2 & 20 & 19$\left(0.072\right)$ & 1.1$\left(0.043\right)$ & 13$\left(0.51\right)$ & steady & polar\tabularnewline
19 & 3 & 20 & 29$\left(0.080\right)$ & 1.7$\left(0.047\right)$ & 31$\left(2.6\right)$ & steady & polar\tabularnewline
20 & 4 & 20 & 39$\left(0.31\right)$ & 2.3$\left(0.02\right)$ & 54$\left(16\right)$ & steady & conic\tabularnewline
21 & 6 & 20 & 58$\left(3.0\right)$ & 3.5$\left(0.2\right)$ & 113$\left(47\right)$ & semi-steady & conic\tabularnewline
22 & 8 & 20 & 77$\left(3.6\right)$ & 4.6$\left(0.2\right)$ & 169$\left(66\right)$ & semi-steady & wide conic\tabularnewline
$\cdots$ & $\cdots$ & $\cdots$ & $\cdots$ & $\cdots$ & $\cdots$ & $\cdots$ & $\cdots$\tabularnewline
23$^{*}$ & 1 & 2 & 8.3$\left(0.38\right)$ & 0.49$\left(0.02\right)$ & 2.0$\left(1.8\right)$ & semi-steady & narrow conic\tabularnewline
24$^{*}$ & 2 & 2 & 16$\left(4.7\right)$ & 0.95$\left(0.28\right)$ & 12$\left(7.0\right)$ & semi-steady & conic\tabularnewline
25$^{*}$ & 4 & 2 & 40$\left(13\right)$ & 2.4$^{**}$$\left(0.77\right)$ & 14$\left(11\right)$ & semi-steady & conic\tabularnewline
26$^{*}$ & 10 & 2 & 140$\left(91\right)$ & 8.3$^{**}$$\left(5.4\right)$ & 23$\left(41\right)$ & semi-steady & conic\tabularnewline
$\cdots$ & $\cdots$ & $\cdots$ & $\cdots$ & $\cdots$ & $\cdots$ & $\cdots$ & $\cdots$\tabularnewline
27$^{\mathrm{\dagger}}$ & 1 & 2 & 4.7$\left(0.29\right)$ & 0.27$\left(0.02\right)$ & 5.4$\left(1.5\right)$ & semi-steady & conic\tabularnewline
\hline
\end{tabular}


\raggedright

({*})~Luminosity is limited to $\Gamma\leq1$ and not completely
self-consistent. 

({*}{*})~$\Gamma$ value corresponds to the luminosity which would
have been achieved in the system is if the luminosity is computed
from the mass accretion rate listed in the previous column. 

($\dagger$)~Fixed luminosity model in \citet{Proga:2008} (their
Run~Cr) and \citet{kurosawa:2008b} (their Model~II). 

($\ddagger$)~Values in brackets are the standard deviations of the
time series values. 

\end{table*}


\section{Results}

\label{sec:Results}

We examine the effect of changing the outer boundary conditions on
the gas property, flow morphology and gas dynamics. Here, we focus
on two key parameters, i.e.~density ($\rho_{\mathrm{o}}$) and temperature
($T_{\mathrm{o}}$) at $r=r_{\mathrm{o}}$. We consider three different
temperatures: $T_{\mathrm{o}}=2\times10^{6}$, $2\times10^{7}$ and
$2\times10^{8}\,\Kelvin$ with a wide range of $\rho_{\mathrm{o}}$.
In total, 22 combinations of $\rho_{\mathrm{o}}$ and $T_{\mathrm{o}}$
are considered. The model parameters along basic results are summarised
in Table.~\ref{tab:Model-Summary}.

\subsection{Dependency of Mass Outflow Rates on Accretion Luminosity}
\label{sub:Mdot-L-relation}


\begin{figure}

\begin{center}
\includegraphics[clip,width=0.48\textwidth]{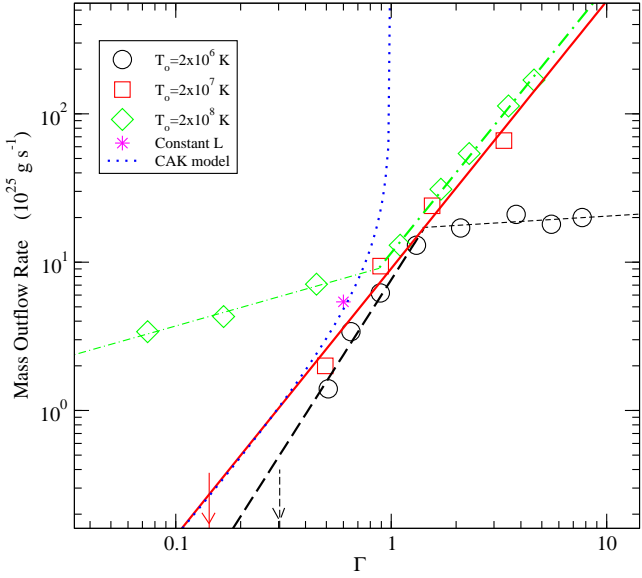}
\par\end{center}

\caption{Comparisons of the relations between the mass outflow rates
($\dot{M}_{\mathrm{out}}$ through the outer boundary) and the Eddington
ratio ($\Gamma$) for the models with
three different ambient gas temperatures ($T_{\mathrm{o}}$) : $2\times10^{6}$
(\emph{circles}), $2\times10^{7}$ (\emph{squires}) and $2\times10^{8}\,\Kelvin$
(\emph{diamonds}). The parameters used for the models/data points
are summarised in Table~\ref{tab:Model-Summary}. The points along
the diagonal lines (thick lines) clearly shows that $\dot{M}_{\mathrm{out}}$
correlates with $\Gamma$ for all $T_{\mathrm{o}}$. The strong correlation
is seen in the lowest temperature models ($T_{\mathrm{o}}=2\times10^{6}\,\Kelvin$)
for $0.5\lesssim\Gamma\lesssim1.5$, but $\dot{M}_{\mathrm{out}}$
becomes saturated at the higher $\Gamma$ values ($\Gamma\gtrsim1.5$;
cf.~Sec,~\ref{sub:Disk-Wind-Like}). For the models with $T_{\mathrm{o}}=2\times10^{7}\,\Kelvin$,
a correlation is seen for all the points with $\Gamma\gtrsim0.5$.
The high temperature models ($T_{\mathrm{o}}=2\times10^{8}\,\Kelvin$)
show two distinctive regions in which the points follow two different
slopes. The outflows in the models for $\Gamma\lesssim1$ (Models~15,
16 and 17) are \emph{mainly thermally driven}, but the models with
$\Gamma\gtrsim1$ are \emph{mainly radiation-driven}. The thermal
winds found here are artifacts of the inner boundary. A simple power
law is used to fit the points for the models with the outflows mainly driven
by radiation, for each temperature: $2\times10^{6}$ (\emph{thick dashed
line}), $2\times10^{7}$ (\emph{thick solid} \emph{line}) and $2\times10^{8}\,\Kelvin$
(\emph{thick dash-dot line}). A separate power law (\emph{thin dashed
line}) is used for the points with four largest $\Gamma$ (Models~6,
7, 8 and 9) in the $T_{\mathrm{o}}=2\times10^{6}$~K models, and
another power law is used for the models with the thermally driven
outflows with $T_{\mathrm{o}}=$ $2\times10^{8}\,\Kelvin$ (Models~15,
16 and 17). The indices of the power law fits are summarised in Table~\ref{tab:Fit-Summary}.
The mass-loss rate predicted by the radiation-driven stellar wind
model (\emph{dotted} \emph{line}) by \citet{Castor:1975} (CAK model)
is also shown for a comparison. The arrows near the bottom of the
panel indicate the approximate location of the Eddington ratios for
the models below which an outflow does not form. The solid and dashed
arrows are for the models with $T_{\mathrm{o}}=2\times10^{6}$ and
$2\times10^{7}$~K, respectively. The constant luminosity ($\Gamma=0.6$)
model (\emph{star}), as in Papers~II and III (with $T_{\mathrm{o}}=2\times10^{7}\,\Kelvin$
and $\rho_{\mathrm{o}}=1\times10^{-21}\,\mathrm{g\, cm^{-3}}$), is
also shown for a comparison (Model~27). }

\label{fig:MdotOut_Gamma}

\end{figure}



\begin{table}

\caption{ \label{tab:Fit-Summary}Summary of power-law fits for the
$\dot{M}_{\mathrm{out}}$-- $\Gamma$ relations}

\begin{tabular}{lccc}
\hline 
Main wind driving force & $T$ & $\Gamma$ range & $q^{**}$\tabularnewline
 & $\left(10^{7}\,\mathrm{K}\right)$ &  & \tabularnewline
\hline 
Radiation  & all$^{*}$ & $\sim0.5$ to $\sim5$ & $2.0\left(\pm0.1\right)$\tabularnewline
Radiation  & $0.2$ & $\sim0.5$ to $\sim1.4$ & $2.3\left(\pm0.3\right)$\tabularnewline
Radiation & $2$ & $\sim0.5$ to $\sim3$ & $1.8\left(\pm0.2\right)$\tabularnewline
Radiation & $20$ & $\sim0.9$ to $\sim5$ & $1.8\left(\pm0.1\right)$\tabularnewline
\ldots{} & \ldots{} & \ldots{} & \ldots{}\tabularnewline
Thermal  & $20$ & $\lesssim0.7$  & $0.4\left(\pm0.1\right)$\tabularnewline
Radiation (`disc-wind') & $0.2$ & $\gtrsim1.4$ & $0.1\left(\pm0.1\right)$\tabularnewline
\hline
\end{tabular}

({*})~Used the points from all three temperatures (excluding points
for Models~6, 7, 8, 9, 15, 16 and 17)

({*}{*})~The index of the power-law, i.e.~$\dot{M}_{\mathrm{out}}\propto\Gamma^{q}$.
The radiation-driven stellar wind model of \citet{Castor:1975} finds
$q=1.67$. 

\end{table}


First, we examine how mass outflow rates of systems depend on the accretion
luminosity that is self-consistently determined. Fig.~\ref{fig:MdotOut_Gamma}
shows the mass outflow rate at the outer boundary $\dot{M}_{\mathrm{out}}\left(r_{\mathrm{o}}\right)$
plotted as a function of the accretion luminosity, or equivalently as a
function of the Eddington ratio $\Gamma$ ($\equiv L/L_{\mathrm{Edd}}$
where $L_{\mathrm{Edd}}=4\pi cGM_{\mathrm{BH}}/\sigma_{e}$ ) for
different combinations of $\rho_{\mathrm{o}}$ and $T_{\mathrm{o}}$.
The actual values for $\dot{M}_{\mathrm{out}}\left(r_{\mathrm{o}}\right)$
and $\Gamma$ (along with the corresponding mass inflow fluxes at the
inner boundary $\dot{M}_{\mathrm{in}}$)
are also listed in Table.~\ref{tab:Model-Summary}. Note that $\dot{M}_{\mathrm{out}}$
values used here are the time-averaged values since the mass inflow
rate $\dot{M}_{\mathrm{in}}$ or equivalently mass accretion rate
$\dot{M}_{\mathrm{a}}$ in some of the models (as also indicated in
the table) show some degree of variability, which will be discussed
later in Section~\ref{sub:Time-Dependent-Behaviours}. 

The figure shows that most of the data points follow the lines that
cross the panel almost diagonally, displaying strong correlations
between $\dot{M}_{\mathrm{out}}$ and $\Gamma$. 
The outflows formed in the models with $\Gamma \gtrsim 0.5$ are mainly \emph{radiatively
driven} whereas those in the highest temperature cases
($T_{\mathrm{o}}=2\times10^{8}\,\Kelvin$) with $\Gamma \lesssim 0.5$
(Models~15, 16 and 17) are mainly \emph{thermally driven}. 

The models with the lowest temperature ($T_{\mathrm{o}}=2\times10^{6}\,\Kelvin$)
follow the main trend defined by the data points for the models with
the radiation-driving dominated outflows for $0.5\lesssim\Gamma\lesssim1.5$.
No outflow forms for $\Gamma\lesssim0.5$. The mass outflow rates
become saturated at higher $\Gamma$ values ($\Gamma\gtrsim1.5$),
i.e.~the strength of the correlation weakens dramatically. The cause
of the saturation seen in the the low temperature models will be discussed
later in Section~\ref{sub:Disk-Wind-Like}. The models with $T_{\mathrm{o}}=2\times10^{7}\,\Kelvin$
shows a correlation between $\dot{M}_{\mathrm{out}}$ and $\Gamma$
for all models with $\Gamma\gtrsim0.5$. Again no outflow forms
for $\Gamma\lesssim0.5$. The figure also shows a point for the model
run (Model~27) with the same parameters as in Model~11
($\mathrm{\rho_{\mathrm{o}}=1\times10^{-21}\,\mathrm{g\, cm^{-3}}}$ 
and $T_{\mathrm{o}}=2\times10^{7}\,\mathrm{K}$), but with a fixed
accretion luminosity ($\Gamma=0.6$). Note that this model is equivalent
to Run~Cr in Paper~II, and Model~II in Paper~III. These fixed
$\Gamma$ (constant $L$) models are located near the main trend of
the $\dot{M}_{\mathrm{out}}$ -- $\Gamma$ relation found here. This
reassures that our previous models parameters ($\Gamma,$ $\rho_{\mathrm{o}}$
and $T_{\mathrm{o}}$) were reasonable, and were somewhat consistent
with each others. The high temperature models ($T_{\mathrm{o}}=2\times10^{8}\,\Kelvin$)
show two distinctive regions in which the $\dot{M}_{\mathrm{out}}$ --
$\Gamma$ relation follows two different
slopes. As mentioned earlier, the outflows found in the models with
$\Gamma\lesssim1$ (Models 15, 16 and 17) are mainly thermally driven,
but the models with $\Gamma\gtrsim1$ are mainly radiation-driven.
The change in the main wind driving mechanism (from thermal pressure
to radiation pressure) is the cause of the change in the strength
or the slope of the $\dot{M}_{\mathrm{out}}$ -- $\Gamma$ correlation.
The cause of the super-Eddington accretion luminosity ($\Gamma>1$)
will be discussed later in Section~\ref{sub:Accretion-with-Super-Eddington}.

A simple power law ($\dot{M}_{\mathrm{out}}\propto\Gamma^{q}$) is
use to fit the points for the models with radiation-driving dominated
outflows, separately for each temperature. We find the power law indices
($q$) as $2.3\left(\pm0.3\right)$, $1.8\left(\pm0.2\right)$ and
$1.8\left(\pm0.1\right)$ for $T_{\mathrm{o}}=$ $2\times10^{6}$,
$2\times10^{7}$ and $2\times10^{8}\,\Kelvin$, respectively. These 
values are very similar to each others, i.e. the slope of the
correlation is independent of or very insensitive to
the outer boundary temperature.  
The slope for $T_{\mathrm{o}}=$ 
$2\times10^{6}$~K seems to be slightly higher than those of the
higher temperature models; however, when the point corresponding to Model~2
($\rho_{\mathrm{o}}=2\times10^{-21}\,\mathrm{g\, cm^{-3}}$) 
is excluded from the line fit, we obtain the power law index $q=1.9$
which is in agreement with those of the models with the higher temperatures.
We note that the outflow in Model~2 is relatively weak and it may
not be entirely radiation-driven as its $\Gamma$ is very close to
that of Model~1, which does not produce an outflow. 

The figure also shows the mass-loss rate predicted by the radiation-driven
wind model of \citet{Castor:1975} (CAK model) for a comparison. The
CAK model predicts $q=1.67$ for $\Gamma\ll1$ and $q\geq 2$ 
for $\Gamma\simless 1$ which are very similar to the slopes found here. 
Their model assumes spherical symmetry,
and the mass outflow rates becomes infinity as $\Gamma$ approaches
to 1. The differences between their model and our model are not only
due to the assumed symmetries (spherical versus axial), but also in
the implementation of the underlining luminosities. Our wind driving
luminosity is coupled to the mass inflow/accretion rate of the system
while their luminosity is fixed constant and they do not have inflows
in their model.

Motivated by the insensitivity of the slope of $\dot{M}_{\mathrm{out}}$
-- $\Gamma$ curves on $T_{\mathrm{o}}$, we compute the slope using
the points from all three temperatures (excluding the thermally-driving
dominated models and the lowest temperature models with $\Gamma \gtrsim
1.5$: Models~6, 7, 8, 9, 15, 16 and 17). The corresponding slope
is found as $2.0\left(\pm0.1\right)$. A separate power law is applied
to the points with four highest $\Gamma$ (Models~6, 7, 8 and 9)
in the $T_{\mathrm{o}}=2\times10^{6}$~K models, and its index is
 $0.1\left(\pm0.1\right)$. For the models with the thermally
driven outflows with $T_{\mathrm{o}}=$ $2\times10^{8}\,\Kelvin$
(Models~15, 16 and 17), the power law index is $0.4\left(\pm0.1\right)$.
Table~\ref{tab:Fit-Summary} summarises the power law indices found
in this analysis. We note that theoretical and numerical studies of
radiation driven stellar and disc winds (e.g.~\citealt{Castor:1975};
\citealt*{proga:1998}; \citealt*{Proga:1999}) predict $q\geq1.67$, which
is quite close to the values found in this study. 

When the outer boundary density $\rho_{\mathrm{o}}$ becomes too small
($\rho_{\mathrm{o}}\lesssim1\times10^{-21}\,\mathrm{g\, cm^{-3}}$),
as in Model~1, no outflow is formed for
$T_{\mathrm{o}}=2\times10^{6}$~K. Similarly, no outflow is found for
the models with $\rho_{\mathrm{o}}\lesssim4\times10^{-22}\,\mathrm{g\, cm^{-3}}$ 
as in Model~10, for $T_{\mathrm{o}}=2\times10^{7}$~K. Unlike
in the lower temperature models, an outflow forms
for $T_{\mathrm{o}}=2\times10^{8}$~K, even with a relatively
small outer boundary density ($\rho_{\mathrm{o}}=0.5\times10^{-22}\,\mathrm{g\, cm^{-3}}$).
The minimum outer boundary density $\rho_{\mathrm{o}}^{\mathrm{min}}$
or the minimum Eddington ratio $\Gamma^{\mathrm{min}}$ above which
an outflow can be formed depends on the ambient temperature ($T_{\mathrm{o}}$).
The higher the value of $T_{\mathrm{o}}$, the smaller the value of
$\Gamma^{\mathrm{min}}$. For example, $\Gamma^{\mathrm{min}}\sim0.3$,
$\sim0.2$ and $<0.05$ for $T_{\mathrm{o}}=$ $2\times10^{6}$, $2\times10^{7}$
and $2\times10^{8}\,\Kelvin$, respectively. When the outer boundary
temperature becomes higher, the average temperature of the gas in
the temperature becomes also higher. This makes the gas to be driven
thermally more easily; hence, the outflow starts to form even at lower
value of $\Gamma$. Note that the Eddington ratio $\Gamma$, by definition,
only considers the radiation force due to Thomson scattering. The
radiation force considered here includes also line scattering process
(see e.g.~Paper~I for our implementation of the radiation force);
hence, the total radiation force could exceed the gravity even for
$\Gamma<1$. 

The correlation between $\dot{M}_{\mathrm{out}}$ and $\Gamma$ (for
the radiation driven outflow cases) may continue for higher
luminosities (i.e.~$\Gamma > 5$), but we are not be able to confirm this
due to a limitation of our code. Strong shocks form in the flows in the models with $\Gamma>5$
(for $T_{\mathrm{o}}=2\times10^{7}$ and $2\times10^{8}$~K models),
and the code fails to handle them.  Unfortunately, this prevents us from exploring
models with very high accretion luminosities.

\subsection{Time-Dependent Behaviours of Mass-Accretion Rates}
\label{sub:Time-Dependent-Behaviours}

\begin{figure}

\begin{center}

\includegraphics[clip,width=0.45\textwidth]{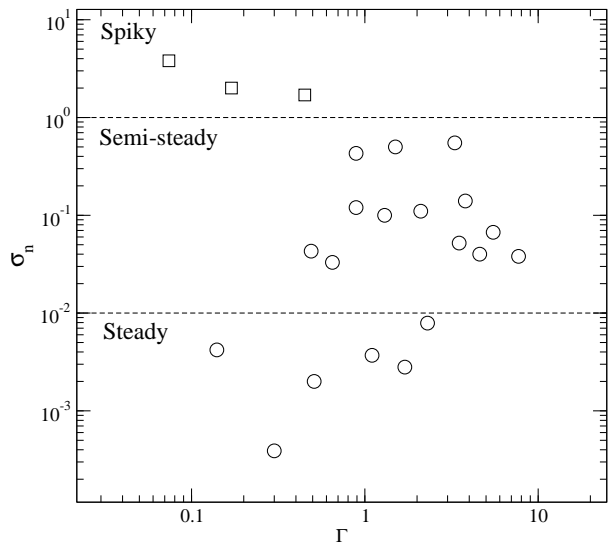}

\end{center}

\caption{Distribution of the normalised standard 
deviation  $\sigma_{n}$ of the mass accretion rates computed at the
inner boundary for Models~1--22 in Table~\ref{tab:Model-Summary} is
shown as a function of the Eddington ration $\Gamma$. 
models are categorised as \emph{steady}, \emph{semi-steady} or
\emph{spiky} when $\sigma_{n}<0.01$, $0.01<\sigma_{n}<1.0$ and
$\sigma_{n}>1.0$, respectively.  The boundaries between different
types are indicated by horizontal lines (\emph{dashed line}). The models
which belong to the \emph{steady} and \emph{semi-steady} types are
indicated by \emph{circles}, and those belong to the \emph{spiky} are
indicated by \emph{squares}. A very weak correlation between
$\sigma_{n}$ and $\Gamma$ is seen for the data points belong to the
\emph{steady} and \emph{semi-steady} types.} 

\label{fig:mdot_var_deine}

\end{figure}

Not all the models we consider here reach steady state solutions and
have smooth inflows (and outflows). Some models do show variability
in the mass-accretion rate ($\dot{M}_{\mathrm{a}}$) (the mass flux
across the inner boundary of our computational domain). 
\textcolor{black}{
To quantify the degrees of variability, we compute the standard
deviations of the time series values of $\dot{M}_{\mathrm{a}}$ and
$\Gamma$ for each model. The results are shown in
Table~\ref{tab:Model-Summary}.  Further, to compare the amount of
variability from different models more directly, we normalise the
standard deviation by the time-averaged values of
$\dot{M}_{\mathrm{a}}$. In other words, we compute the ratio
($\sigma_{n}$) of the standard deviation of $\dot{M}_{\mathrm{a}}$ to
the time-averaged values of $\dot{M}_{\mathrm{a}}$.  The distribution
of the normalised standard deviation $\sigma_{n}$ is shown in
Fig.~\ref{fig:mdot_var_deine}.  The figure shows $\sigma_{n}$ as a
function of $\Gamma$ for Models~1--22 in
Table~\ref{tab:Model-Summary}.
}

\begin{figure}

\begin{center}

\includegraphics[clip,width=0.45\textwidth]{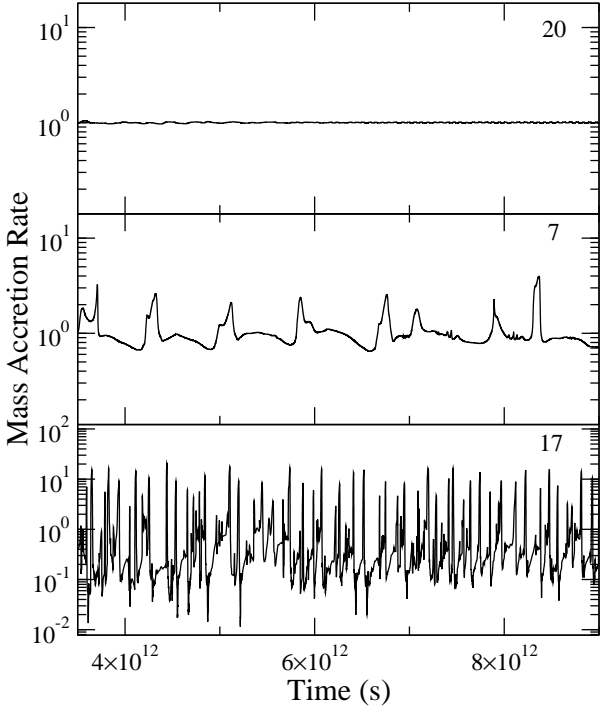}

\end{center}

\caption{Examples of different types of time-dependent mass-accretion
rate variability patterns. The panels show the mass-accretion rates
as a function of time for selected samples of models which show steady
(top), semi-steady (middle) and spiky (bottom) variabilities (see
Table~\ref{tab:Model-Summary}). The number on the upper right hand
corner of each panel indicates the corresponding model number in Table~\ref{tab:Model-Summary}.
The mass accretion rates on the vertical axes are in units of the
time-averaged mass accretion rate of each model which are 39, 15 and
7.5$\times10^{25}\,\mathrm{g\, s^{-1}}$ from the top
to bottom panels, respectively.}

\label{fig:mdot_var}

\end{figure}

\textcolor{black}{
We categorise the time-dependent natures of $\dot{M}_{\mathrm{a}}$ in
our simulations into three types: (1)~steady, (2)~semi-steady and
(3)~spiky, based on the value $\sigma_{n}$. We define a model as
\emph{steady} when $\sigma_{n}<0.01$, and as \emph{semi-steady} when
$0.01<\sigma_{n}<1.0$. Further, it is defined as  
\emph{spiky} when $\sigma_{n}>1.0$.  Based on this definition, the
variability type of each model is assigned and placed also in
Table~\ref{tab:Model-Summary}. 
Out of 22 models considered here, 6 are found as 
\emph{steady}, and 13 as \emph{semi-steady}. 
Only 3 models are found as  \emph{spiky}. 
Fig.~\ref{fig:mdot_var} shows examples of the three different types 
of $\dot{M}_{\mathrm{a}}$ variability found in our simulations.  
}

In the models with steady $\dot{M}_{\mathrm{a}}$, inflows and outflows
are smooth and the level of $\dot{M}_{\mathrm{a}}$ variability (the
change with respect to the time averaged value of $\dot{M}_{\mathrm{a}}$)
is typically less than 1~per~cent. In the models with semi-steady
$\dot{M}_{\mathrm{a}}$, small-sized and relatively high density structures
are occasionally formed in the flows, and some of them fall toward
the centre. This causes moderate levels of variability ($\lesssim$200~per~cent)
in a relatively short timescale ($\sim10^{11}$~s). The models marked
as \emph{spiky} occurs only in the high temperature models ($T_{\mathrm{o}}=2\times10^{8}\,\Kelvin$),
and in the cases when the outflows is thermally driven. The relatively
high temperature of gas keeps the sound speed to be high.
As a result, the outflows in these models are sub-sonic. The effect
of a small density perturbation in the flow that occurs near the inner
boundary propagates though the entire computational space in relatively
small timescale (because of the high sound speed) compared to the
dynamical timescale of the flows. The acoustic wave generated near
inner boundary propagates outward and bounces back at
$r=r_{\mathrm{o}}$ in the hot media, when the outflow is relatively weak and when the
radiative force is relatively weak. This causes rather stochastic
sharp peaks and dips (spikes) in the $\dot{M}_{\mathrm{a}}$ variability
curves. The level of variability is rather large ($>100$~per~cent)
in these models. We do not consider the models with spiky $\dot{M}_{\mathrm{a}}$
(i.e.~Models~15, 16 and 17) as physical because they are affected
by the inner boundary. The inner boundary is too far from the BH for
the inflow to become supersonic in those cases; hence, it can reflect the
flows and turn inflows into outflows. 

\textcolor{black}{
When we exclude the unphysical spiky $\dot{M}_{\mathrm{a}}$ models
(i.e.~Models~15, 16 and 17) from Fig.~\ref{fig:mdot_var_deine}, we
find a very weak correlation between $\sigma_{n}$ and $\Gamma$, 
i.e.~the larger the accretion luminosity, the amount of variability
tends to be larger. The distribution of $\sigma_{n}$ seems to be 
continuous (as a function $\Gamma$). There is no clear separation of
the two populations: the \emph{steady} and \emph{semi-steady} types, 
according to the figure. 
}


\subsection{Disc Wind Like Solutions}
\label{sub:Disk-Wind-Like}

In most of the models explored (Table~\ref{tab:Model-Summary}),
the flow morphology are quite similar to the ones presented earlier
(Papers~I, II and III) e.g.~bipolar outflows and fragmented dense
cold cloud outflows. Here we report a new morphological type of solutions
which has not been seen in our previous models. Among the low temperature
($T_{\mathrm{o}}=2\times10^{6}$) models, when the density at
$r=r_{\mathrm{o}}$ is relatively high ($\rho_{\mathrm{o}}>5.0\times10^{-20}\,\mathrm{g\, cm^{-3}}$),
the flow starts to resemble a radiation driven disc wind which arises
from the very inner part of accretion discs
(e.g.~\citealt{proga:1998}; \citealt{proga:1999b}),  
 \textcolor{black}{
 and a thermally-driven disc wind which arises from
relatively large radii (e.g.~\citealt{Woods:1996}; \citealt{Proga:2002b}). 
}

\begin{figure*}

\begin{center}
\begin{tabular}{cc}
\includegraphics[clip,height=0.27\textheight]{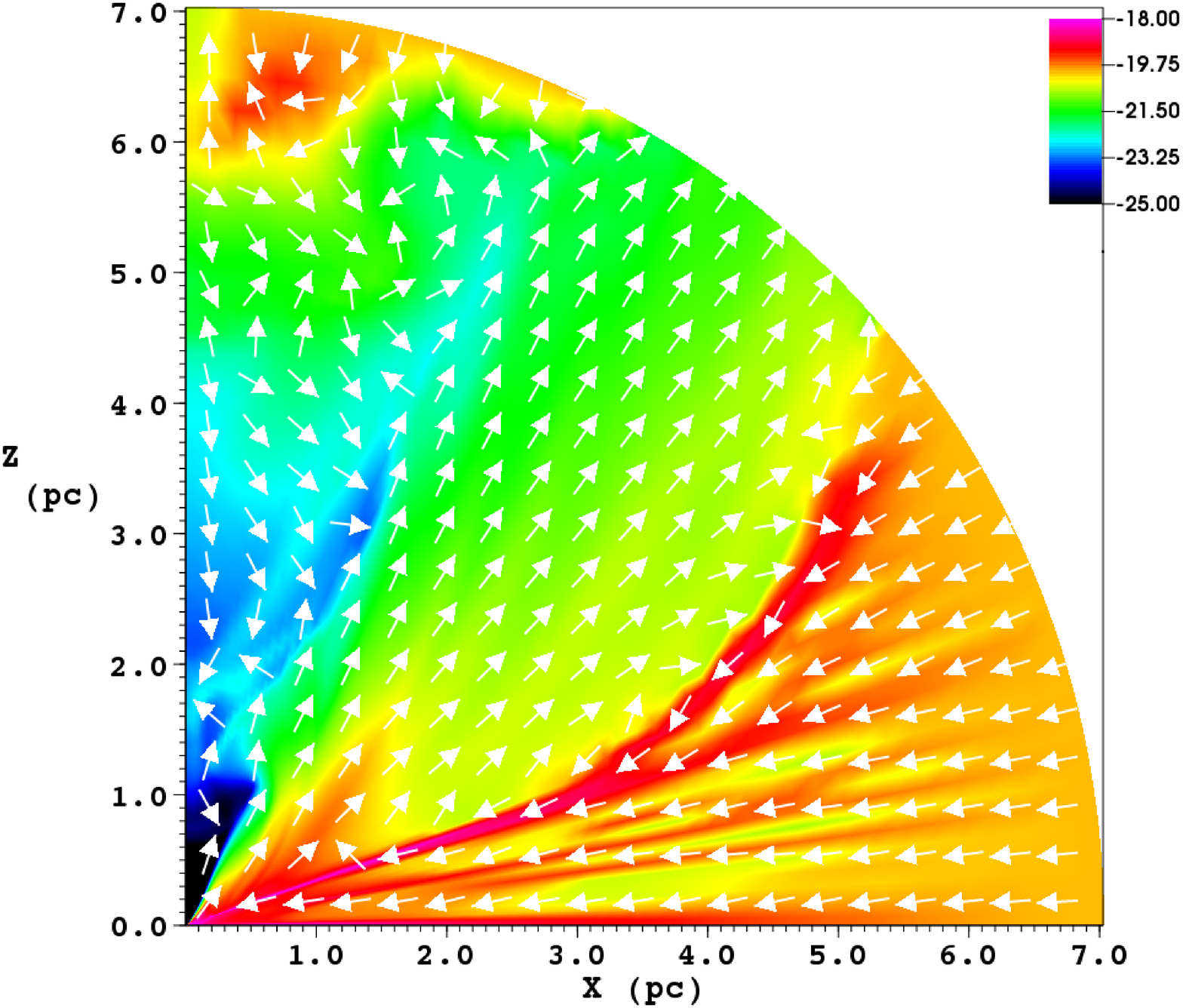} & \includegraphics[clip,height=0.27\textheight]{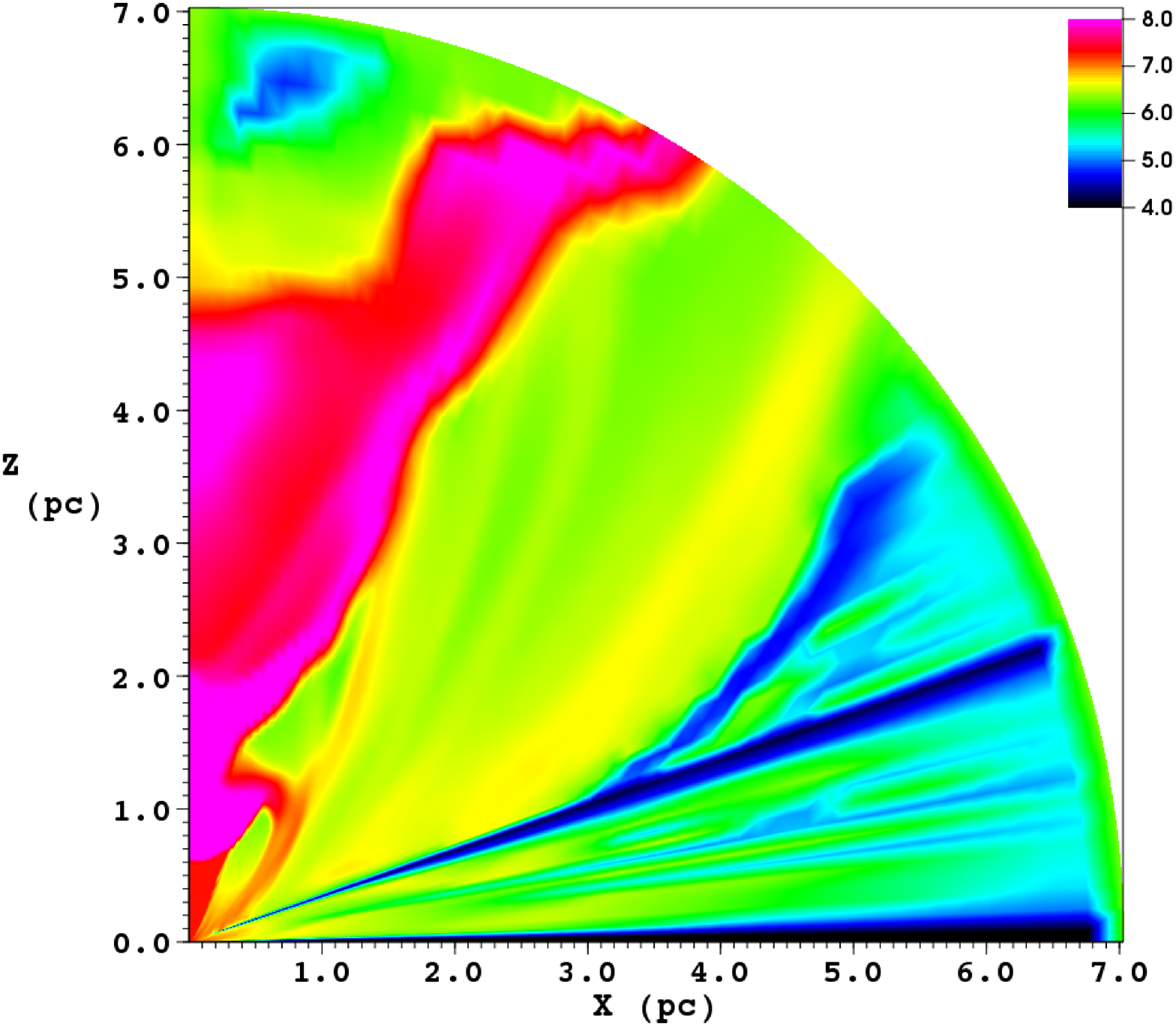}\tabularnewline
\includegraphics[clip,height=0.27\textheight]{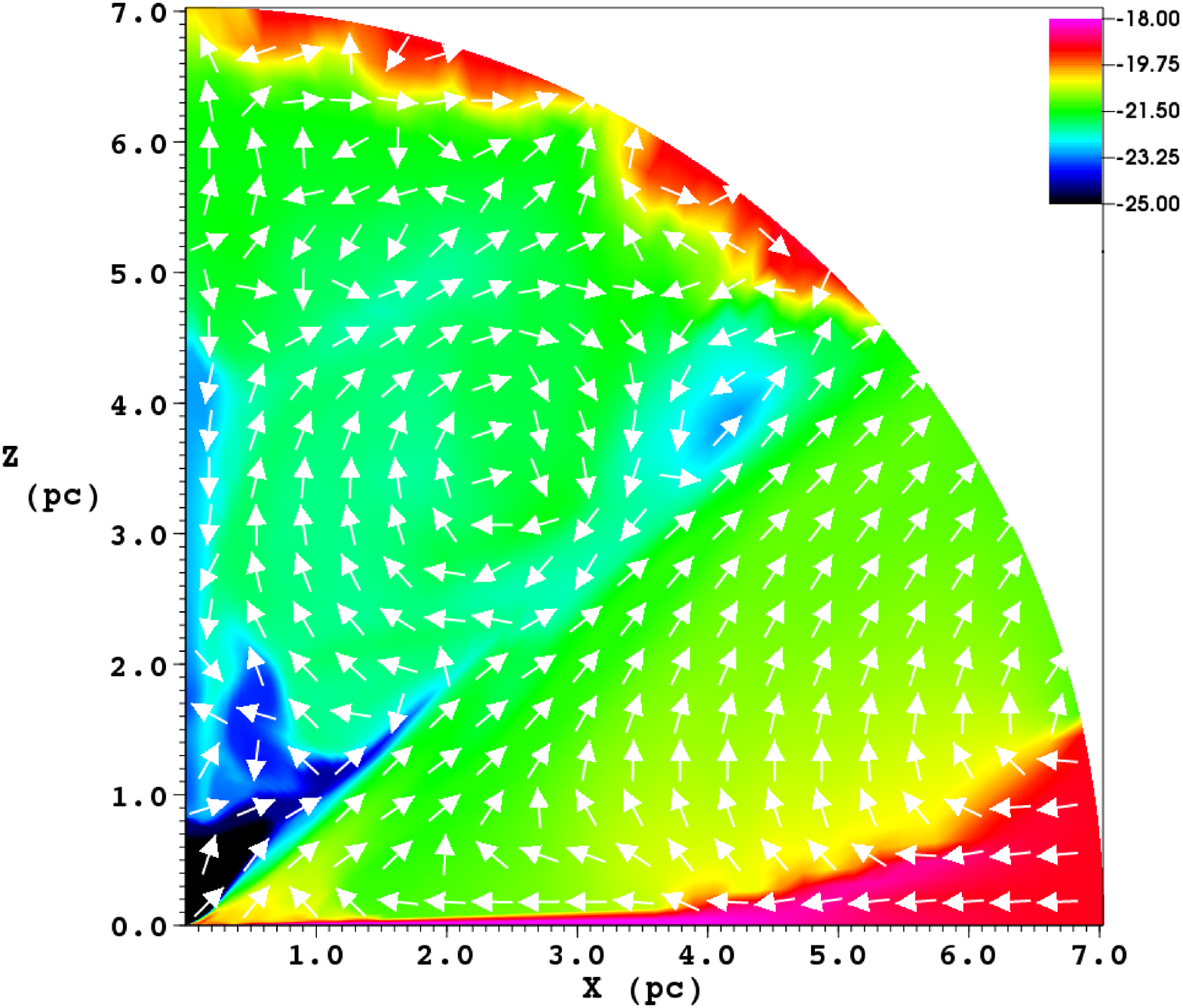} & \includegraphics[clip,height=0.27\textheight]{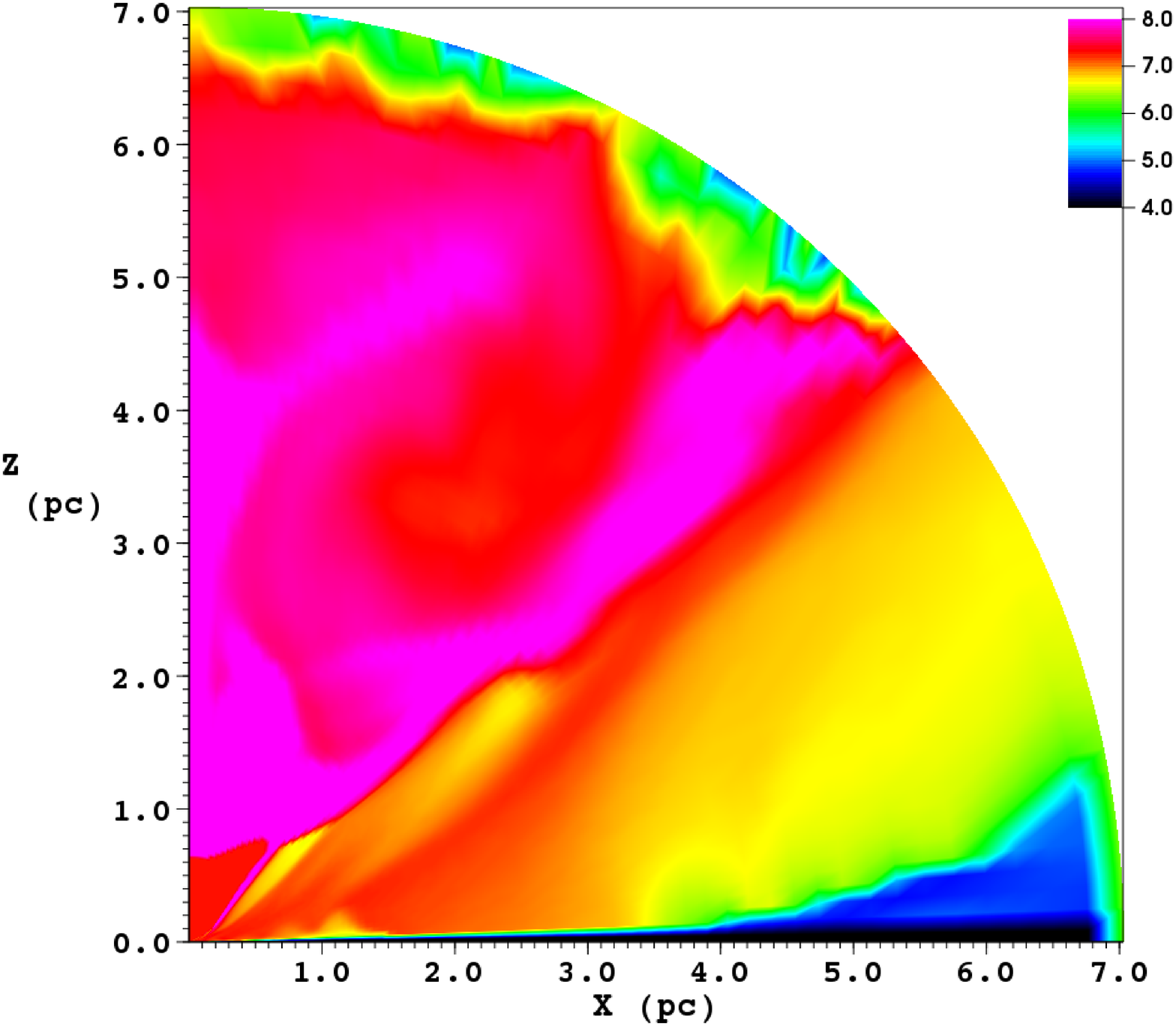}\tabularnewline
\includegraphics[clip,height=0.27\textheight]{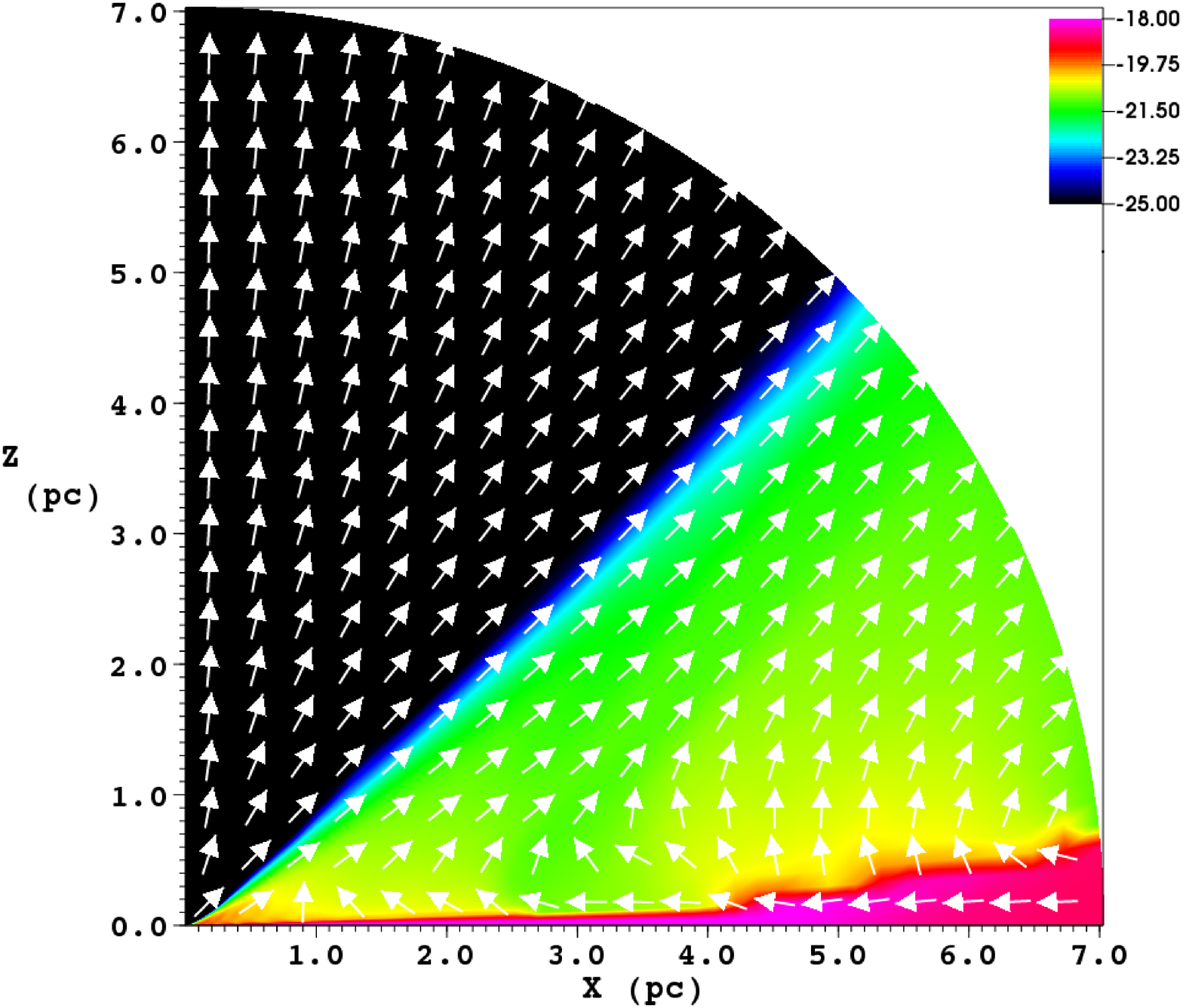} & \includegraphics[clip,height=0.27\textheight]{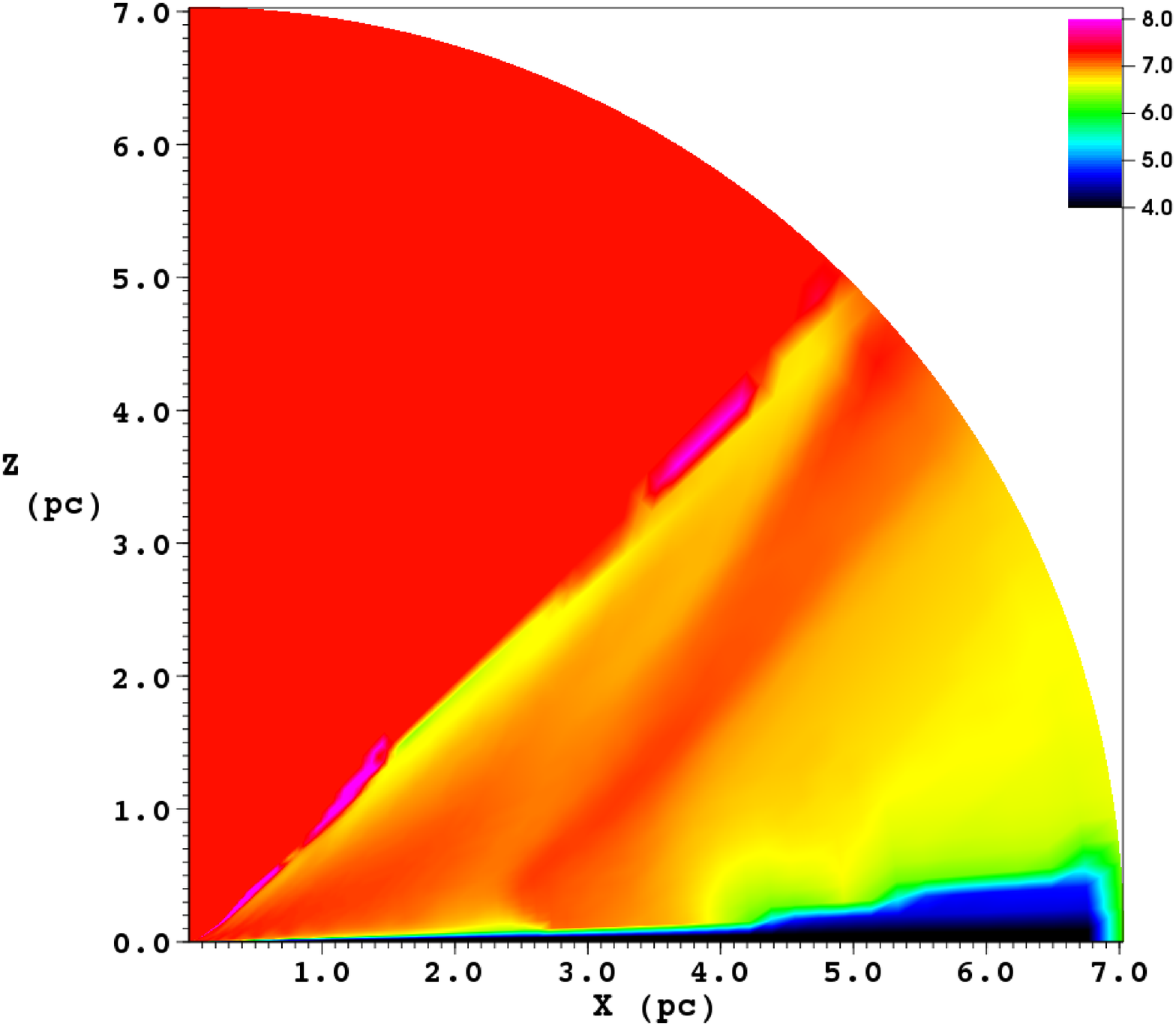}\tabularnewline
\end{tabular}
\par\end{center}

\caption{A transition to a disc-wind like solution in the low temperature
models ($T_{\mathrm{o}}=2\times10^{6}\,\mathrm{K}$). The density
(in logarithmic scale) over-plotted with the directions of poloidal
velocity as arrows (\emph{left column}) for Models~5 (\emph{top}),
8 (\emph{middle}) and 9 (\emph{bottom}) are shown. The figures are
placed in order of increasing density at $r=r_{\mathrm{o}}$ ($\rho_{\mathrm{o}}$),
from the top to bottom (cf.~Table~\ref{tab:Model-Summary}). The
corresponding temperature maps (in logarithmic scale) of each model
are placed in the \emph{right column}. The colour tables for the
density and temperature maps are labelled with the logarithmic values
in $\mathrm{g\, cm^{-3}}$ and K, respectively.  The top density
plot shows the high density gas streams falling toward the centre
in the wide range of polar angles ($\theta\gtrsim50^{\circ}$), and
the outflow is formed only in the mid polar angles $30^{\circ}\lesssim\theta\lesssim50^{\circ}$.
The middle density plot shows the infalling dense gas forming a relatively
thick disc-like structure, and the outflow occurs in larger and wider
polar angles ($45^{\circ}\lesssim\theta\lesssim80^{\circ}$) which
resembles a disc wind (e.g.~\citealt{proga:1998}). The high density
gas near $r=r_{\mathrm{o}}$ at lower polar angles ($\theta\lesssim45^{\circ}$)
prevents the gas from forming outflows in the polar direction. The
bottom density plot shows the accretion region is now squeezed to
a much thinner equatorial disc, and a very wide outflows ($45^{\circ}\lesssim\theta\lesssim85^{\circ}$)
or a disc wind is formed.}

\label{fig:disk-wind-examples}

\end{figure*}

Fig.~\ref{fig:disk-wind-examples} demonstrates how the wind configuration
transforms to a disc wind like solution. For this purpose, we compare
the gas flows from Models~5, 8 and 9 which have the same outer boundary
temperature $T_{\mathrm{o}}=2\times10^{6}\,\Kelvin$ but have different
values of $\rho_{\mathrm{o}}$, i.e.~$1\times10^{-20}$, $1\times10^{-19}$
and $2\times10^{-19}\,\mathrm{g\, cm^{-3}}$ , respectively. In Model~5,
which has the lowest $\rho_{\mathrm{o}}$ value among the three, the
high density gas streams falling toward the centre in relatively wide
range of polar angles ($50^{\circ}\lesssim \theta < 90^{\circ}$). The outflow
is formed only in the mid polar angles $30^{\circ} \lesssim \theta
\lesssim 50^{\circ}$.
When $\rho_{\mathrm{o}}$ is increased 10 times (Model~8), the infalling
dense streams of gas seen in the previous model are converged toward
equatorial plane, and form a relatively dense disc-like structure.
The outflow now occurs at larger polar angles and in wider angle range
($45^{\circ} \lesssim \theta \lesssim 80^{\circ}$). The outflow resembles a
disc wind. However, the high density gas near $r=r_{\mathrm{o}}$ at
lower polar angles $\theta \lesssim 45^{\circ}$ prevents the gas from
forming outflows and keeps the gas relatively 
hot in the polar direction. In this case, no fast low-density polar
wind is formed unlike the ones seen in the models of
\citet{proga:1998}. This difference is caused by the difference in the
treatment of boundary conditions, i.e. in  \citet{proga:1998}, the
density and temperature at $r=r_{\mathrm{o}}$  are unconstrained. 
When we increase the density $\rho_{\mathrm{o}}$ by a factor of 2
(Model~9), the radiative force becomes strong enough to push away the
high density gas in the polar direction, and the fast low-density
polar wind is formed. However, in this case, the outflowing mass-flux
in the polar direction is negligible (due to its very low density) compared 
to the outflow in the higher polar angles ($\theta \gtrsim 45^{\circ}$). The accretion region is
now converged to a much thinner equatorial disc, and a very wide outflow
($45^{\circ} \lesssim \theta \lesssim 85^{\circ}$) or a disc wind is formed. 

The saturation of the mass outflow rates at the high $\Gamma$ values
for $T_{\mathrm{o}}=2\times10^{6}$~K models seen in Fig.~\ref{fig:MdotOut_Gamma}
(Section~\ref{sub:Coupling-of-Mass-Accretion}) can be explained qualitatively
by the change of the outflow pattern to a disc wind like configuration.
Before the saturation occurs, the polar outflow
can become wider by squeezing the high density inflowing stream toward
the equatorial regions. This allows the mass outflow 
rate to grow proportionally with the mass-accretion rate or with the
accretion luminosity $\Gamma$. In the higher $\Gamma$ models,
e.g.~Models~8 and 9, the dense inflowing gas streams are converged all 
the way to the equatorial plane ($\theta\approx90^{\circ}$); hence,
the polar angle range, in which outflow occurs, can not be increased
any more. This limit in the size of the outflow polar angle range
is the cause of the saturation of the mass outflow rates (the flat
part of the $\dot{M}_{\mathrm{a}}$--$\Gamma$ relation in Fig.~\ref{fig:MdotOut_Gamma}).
In addition, once the flow becomes disc-wind like, it is harder for a
system to produce stronger outflows for the following 
reason. In our models, the radiation peaks in polar directions (cf.~$\cos\theta$
dependency of the radiation flux as in Sections~\ref{sub:Overview}
and \ref{sub:Accretion-with-Super-Eddington}); hence, the most of
the radiation escapes in the polar direction. In the disc-wind like
flow morphology, however, there is almost no gas to be blown away
by the radiation in the polar regions (Fig.~\ref{fig:disk-wind-examples}).
This results in the insensitivity of the mass outflow rate with $\Gamma$.
Near the equatorial plane where most of the infalling gas is
concentrated, the radiation is very weak because of  the $\cos{\theta}$
dependency of the disc radiation, and because of the very high optical
depth of the disc like structure.  
Finally, the temperature map for the disc-wind like solution shows
that the temperature of the wind is relatively high ($T\sim10^{7}$~K).
This temperature is too high for the radiation force due to line processes
(line force hereafter) to be effective. We confirmed that the main
wind driving force for the wind in this case is the continuum radiation
force (due to electron scattering). 


\section{Discussions}

\label{sec:Discussions}

\subsection{Accretion with Super-Eddington Luminosity}
\label{sub:Accretion-with-Super-Eddington}

\begin{figure}

\begin{center}
\includegraphics[clip,width=0.48\textwidth]{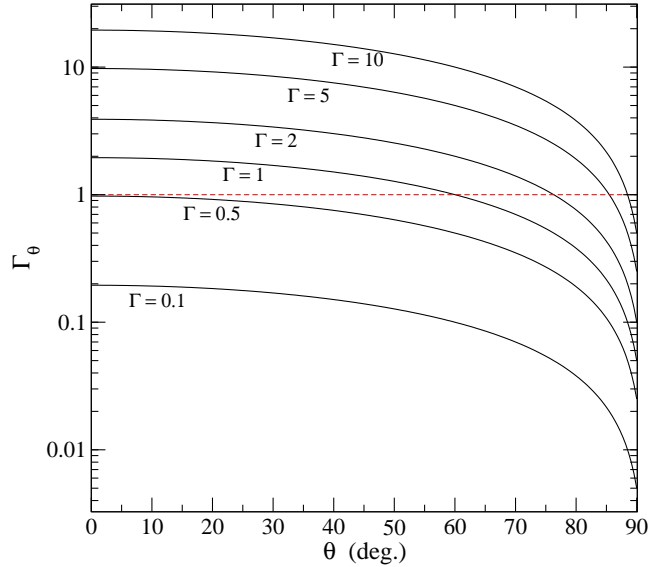}
\par\end{center}

\caption{Ratios ($\Gamma_{\theta}$) of the radiation force due to
electron scattering ($g_{\mathrm{rad}}^{e}$) to the gravitational
force ($g$) as a function of spatial polar angle
$\theta$ for different values of the Eddington ratios $\Gamma$ (\emph{solid
lines}). The lines indicate the lower limit of the ratio because the
radiation force due to line process are not included here. The horizontal
\emph{dashed line} indicates $\Gamma_{\theta}=1$ at which the magnitudes
of radiation force and the gravitational force are equal. In the polar
angle ranges below the $\Gamma_{\theta}=1$ line, an inflow of gas
is possible. This range decreases as $\Gamma$ becomes larger, indicating
that an inflow can occur in larger polar angles, i.e.~closer to the
equatorial planes. Note that even for $\Gamma=5$, $\Gamma_{\theta}$
is still smaller than 1 for $\theta\gtrsim85^{\circ}$, and for $\Gamma=10$
case, $\theta\gtrsim88^{\circ}$. The fractional luminosities of the
spherical X-ray emission and that of the disc UV emission are assumed
to be $0.05$ and $0.95$ respectively (cf.~Section~\ref{sub:Overview}).}

\label{fig:g_rad_vs_g_gravity}

\end{figure}

Many of the self-consistently determined accretion luminosities in
the models presented in Section~\ref{sub:Mdot-L-relation} (see Table~\ref{tab:Model-Summary}
and Fig.~\ref{fig:MdotOut_Gamma}) are found to be above the Eddington
luminosity ($\Gamma>1$). At first, this may be puzzling since one
would expect that the high luminosity hence strong radiation pressure
induced by the high mass accretion rate should regulate the amount
of luminosity and keep $\Gamma<1$. This might be the case if the
gas were not rotating and the radiation were spherically symmetric; however,
the situation changes when the source geometry of the strong UV radiation
is not spherical.  Because of the disc geometry,
the flux of the disc radiation depends on the polar angle. It
radiates more in the polar directions 
(${\cal F}_{{\rm \mathrm{disc}}}\propto|\cos{\theta}|$),
as mentioned in Section~\ref{sub:Hydrodynamics}, and it turns inflows
into outflows. To see how this assumption leads to the super-Eddington
accretion solutions found in our models, consider the ratios ($\Gamma_{\theta}$)
of the radial forces per unit mass (accelerations) by 
radiation ($g_{\mathrm{rad}}$)
and gravity ($g$) (cf.~equation~\ref{eq:hydro02}). To simplify
our argument, we consider only the radiation force due to electron
scattering and exclude that due to spectral line processes here. 
\textcolor{black}{
By setting $\mathcal{M}=0$ in equation~\ref{eq:rad-force-final}, 
the radiative acceleration can be written as 
}
\begin{equation}
g_{\mathrm{rad}}^{e}=\frac{\sigma_{e}\, L}{4\pi\, r^{2}}\left(f_{\mathrm{*}}+2f_{\mathrm{disc}}\cos\theta\right)\label{eq:g_rad_continnum}\end{equation}
where $\sigma_{e}$ and $L$ are the electron scattering cross
section and the total accretion luminosity respectively. The superscript
$e$ on the left term indicates that the acceleration is only due
to electron scattering. The fractional luminosities of the X-ray emitting
spherical region and that of the UV emitting accretion are assumed
to be $f_{\mathrm{*}}=0.05$ and $f_{\mathrm{disc}}=0.95$ (Section~\ref{sub:Overview}).
Using equation~\ref{eq:g_rad_continnum} and the gravitational
acceleration $g=-G\, M_{\mathrm{BH}}/r^{2}$ (assuming it is Newtonian
and a point-mass), the ratio becomes \begin{equation}
\Gamma_{\theta}=\left|\frac{g_{\mathrm{rad}}^{e}}{g}\right|=\Gamma\left(f_{\mathrm{*}}+2f_{\mathrm{disc}}\cos\theta\right)\label{eq:g_rad_over_g_gra}\end{equation}
where $\Gamma=\sigma_{e}L/(4\pi\, GM_{\mathrm{BH}})$ and is the Eddington
ratio. Fig.~\ref{fig:g_rad_vs_g_gravity} shows the values of $\Gamma_{\theta}$
plotted as a function of the polar angle $\theta$ for a range of
the Eddington ratios ($0.1<\Gamma<10$). We consider only the
radiation force due to electron scattering. Therefore, the lines shown in the
figure are only lower limits of $\Gamma_{\theta}$ values
for at a given $\theta$ angle. The figure shows $\Gamma_{\theta}<1$
(the outward radiation force is smaller than the inward gravitation
force) at all $\theta$ for $\Gamma<0.5$ cases. For the models with
$\Gamma>1$, the figure shows that $\Gamma_{\theta}$ can still be
less than 1 at larger values of $\theta$. For example, for $\Gamma=5$
model, $\Gamma_{\theta}$ is still smaller than 1 for $\theta\gtrsim85^{\circ}$,
and for $\Gamma=10$ case, $\theta\gtrsim88^{\circ}$. This critical
angle $\theta_{\mathrm{crit}}$ at which the radiation force equals
the gravity ($\Gamma_{\theta}=1$) becomes lager as the accretion
luminosity or $\Gamma$ increases. The inflow of gas is potentially
possible only in the polar angle range which has $\Gamma_{\theta}<1$.
Therefore,
the location of the accretion streams becomes closer and closer to
the equatorial plane. The most important point of Fig.~\ref{fig:g_rad_vs_g_gravity}
for us is that $\Gamma_{\theta}$ can be still less than 1 near the
equatorial plane which allows an accretion of gas in a super-Eddington
luminosity environment. This is a direct consequence of the disc geometry
which allows most of the radiation to escape in polar direction, but
not in the equatorial direction. In addition, the radiative cooling
in the dense gas confined near the equatorial plane is very efficient,
and the gas temperature is kept low (e.g.~see the bottom panels of
Fig.~\ref{fig:disk-wind-examples}). This also helps to accrete the
matter in the equatorial plane. 

Interestingly, a recent 3-D radiation-hydrodynamic simulation of the
formation of a massive (stellar) binary by accretion by 
\citet{Krumholz:2008} demonstrated that the radiation pressure would not be
able to halt accretion totally even at $\Gamma > 1$  because of
gravitational and Rayleigh Taylor instabilities which channel
the accretion of matter on to the binary from the accretion disc, and 
the self-shielding of radiation by high density gas filaments which
could still experience a net inward force.
\textcolor{black}{
Note that this type of self-shielding does not occur in our model
since the UV radiation, which causes 95~per~cent or more of the total radiation
force in our model, is assumed to be optically thin in our flows.
}



\subsection{Effect of Luminosity Limit }

The self-consistent accretion luminosity models presented in Section~\ref{sub:Mdot-L-relation}
do not limit their luminosities to be below the Eddington luminosity
(i.e.~$\Gamma < 1$) as explained above. A super-Eddington or super-critical
accretion on the SMBH may be possible (e.g.~\citealt{Jaroszynski:1980};
\citealt*{Abramowicz:1980}; \citealt{Begelman:2002}; \citealt{Ohsuga:2002};
\citealt{Ohsuga:2005}; \textcolor{black}{see also a review by
  \citealt{Abramowicz:2005}.}).   
However, we can not be certain whether the
accretion disc would be still stable and would accrete at a such high
Eddington ratio (e.g.~$\Gamma=3.3$ for Model~14) because we do not model
the dynamical evolution of the accretion disc itself (cf.~Section~\ref{sub:Model-Setup}).
\textcolor{black}{
When the accretion disc becomes optically thick, the radiation field
of the interior of the disc itself is expected to become very
isotropic. In general, a super-Eddington accretion in such environment
is much more difficult than the one we considered in this paper. When
the radiation field is isotropic, one would expect the accreting gas
to be turned around when $\Gamma > 1$ and the disc becomes optically
thick; hence, it would tend to keep the mass-accreting rate below the
super-Eddington rate.
}
For this reason, we examine the possibility that the accretion luminosity
is limited to the Eddington luminosity, and study the consequences
of such a limit. 

\begin{figure}

\begin{center}
\includegraphics[clip,width=0.48\textwidth]{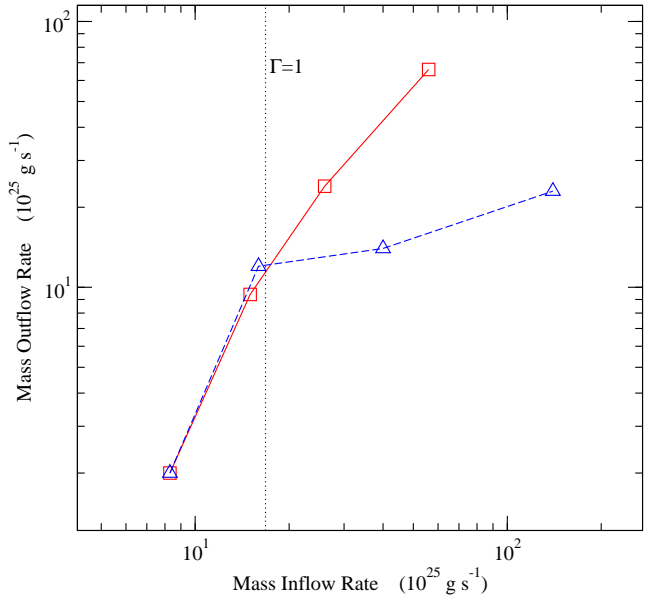}
\par\end{center}

\caption{Effect of limiting accretion luminosity on the mass outflow
rates for the models with the ambient temperature $T_{\mathrm{o}}=2\times10^{7}\,\Kelvin$.
The mass outflow rates are plotted against the mass inflow rates for
the models with (\emph{triangles}) and without (\emph{squares}) the
threshold accretion luminosity for a comparison. For the models with
the luminosity limit, the maximum allowed luminosity is set to the
Eddington luminosity, i.e.~$\Gamma=1$. The mass inflow rate that
corresponds to $\Gamma=1$ is indicated by a vertical \emph{dotted
line}.}

\label{fig:MdotOut_L_limit}

\end{figure}

We recomputed the models with the self-consistent 
evaluation of the accretion luminosity, but now the upper limit 
of the accretion luminosity is
set to the Eddington luminosity. Fig.~\ref{fig:MdotOut_L_limit}
shows the mass outflow rates at $r=r_{\mathrm{o}}$ plotted against
the mass accretion rates at the inner boundary for the models with
$T_{\mathrm{o}}=2\times10^{7}\,\Kelvin$ with and without the upper
limit on the luminosity. Only the models with high mass accretion
rates (near and above the mass accretion rate corresponding to $\Gamma=1$)
are shown in the figure (cf.~Table~\ref{tab:Model-Summary}). The
figure is similar to Fig.~\ref{fig:MdotOut_Gamma}, but the mass
accretion rate is used for the horizontal axis instead of $\Gamma$.
The figure shows that mass outflow weakens (the mass outflow rate
decreases) for the models with larger mass accretion rates (larger than
the value corresponding to $\Gamma=1$). On the other hand, the
models with smaller mass accretion rates (below 
$\Gamma=1$ line in the figure) have nearly identical mass outflow
rates. This behaviour is expected from the luminosity limiting model.
The models with higher mass accretion rates are affected by the luminosity
limit hence have smaller luminosities compared to the cases without
the limit. A smaller luminosity naturally leads to a lower mass
outflow rate. Interestingly this decrease or the saturation 
of the outflow strength (the mass outflow rates) is very similar to
that one seen in the high $\Gamma$ end of the low temperature models
($T_{\mathrm{o}}=2\times10^{6}$~K) in Section~\ref{sub:Mdot-L-relation}
(see Fig.~\ref{fig:MdotOut_Gamma}) although the physical reason
for the cause of the saturation is different. 


\subsection{Effect of unconstraining the outer boundary temperature}
\label{sub:To_not_fixed}

Motivated by the insensitivity of the slopes of
$\dot{M}_{\mathrm{out}}$--$\Gamma$ relations on the outer boundary
temperature ($T_{o}$) found in Section~\ref{sub:Mdot-L-relation}, we now
investigate the effect of allowing the temperature at $r_o$ to be 
computed self-consistently instead of being fixed at some pre-determined
value.
The models presented so far have assumed that the gas
which surrounds our computational domain is `Comptonised' at a constant
temperature $T_{o}$.  Here, we consider 7 additional models in which
we do not constrain the temperatures at $r=r_{\mathrm{o}}$. We now
compute
the temperatures self-consistently by solving the equation of the energy
(i.e., equation~\ref{eq:hydro03}) at $r_o$, as it is  done for any other points in
our computational domain.  The basic model parameters and results are
summarised in Table~\ref{tab:Model-Summary2}.

\begin{figure}

\begin{center}
\includegraphics[clip,width=0.48\textwidth]{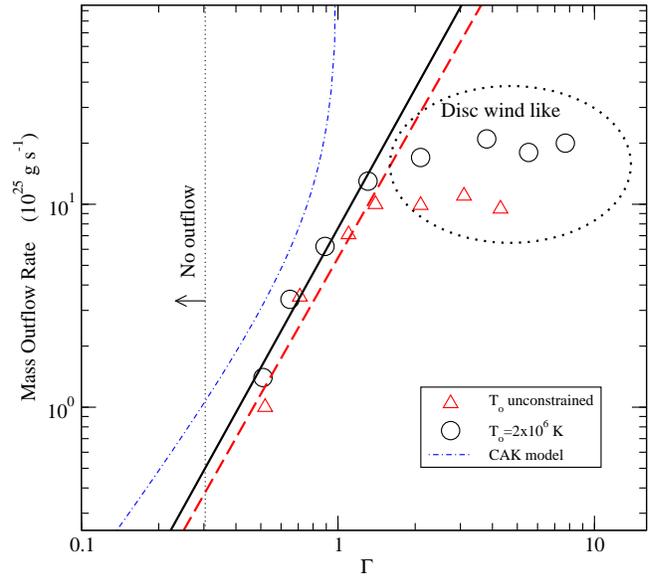}
\end{center}

\caption{Comparison of models with (\emph{circles}) and without
  (\emph{triangles}) fixing the temperature at the outer boundary 
  ($T_{\mathrm{o}}$).  The mass outflow rates 
  ($\dot{M}_{\mathrm{out}}$ through the outer boundary) are plotted as
  a function of the Eddington ration ($\Gamma$) for the models with
  $T_{\mathrm{o}}=2\times 10^6$~K from
  Section~\ref{sub:Mdot-L-relation} (see 
  Table~\ref{tab:Model-Summary}) and for the new models without
 $T_{\mathrm{o}}$ constrained (see Table~\ref{tab:Model-Summary2}).
  Both sets of the models
 show a very similar dependency of $\dot{M}_{\mathrm{out}}$ on
 $\Gamma$. The models shows strong correlations between
 $\dot{M}_{\mathrm{out}}$ and $\Gamma$ when $\Gamma \lesssim 2$. The
 new set of models also show a saturation of $\dot{M}_{\mathrm{out}}$
 beyond $\Gamma \gtrsim 2$ as in the $T_{\mathrm{o}}=2\times 10^6$~K
 (fixed) models, and exhibit the disc wind like solutions (see
 Section~\ref{sub:Disk-Wind-Like}).  A simple power law is used to fit
 the points with $\Gamma \lesssim 2$ (Models~28, 29, 30 and 31 in
 Table~\ref{tab:Model-Summary2}) which show a strong correlation between
 $\dot{M}_{\mathrm{out}}$ and $\Gamma$.  The index of the power law
 fit for the new models is $q=2.2\pm 0.4$ which is very similar to
 that of the $T_{\mathrm{o}}=2\times 10^6$~K models (see
 Table~\ref{tab:Fit-Summary}). 
The mass-loss rate predicted
 by the radiation-driven stellar wind model (CAK model) is also shown
 for a comparison (\emph{dash-dot line}).  The 
 vertical line at $\Gamma\approx0.3$ (\emph{dotted line}) indicates the
 approximate $\Gamma$ value below which a model does not produce
outflows.}

\label{fig:To_test}

\end{figure}

Fig.~\ref{fig:To_test} shows the mass outflow rates
($\dot{M}_{\mathrm{out}}$ through the outer boundary) plotted as a
function of the Eddington ration ($\Gamma$) for the new runs, along
with those of the models with $T_{\mathrm{o}}=2\times 10^6$~K from
Section~\ref{sub:Mdot-L-relation} (see also Table~\ref{tab:Model-Summary}). 
Both sets of the models show a very similar dependency of 
$\dot{M}_{\mathrm{out}}$ on $\Gamma$.  This is mainly because 
the average temperatures (self-consistently determined) at $r_o$
for the new
models considered here are relatively low and similar to that of the fixed outer
boundary temperature models i.e. $T_{\mathrm{o}}=2\times 10^6$~K.
Table~\ref{tab:Model-Summary2} lists the average outer boundary
temperatures for the individual runs.

A strong correlation between $\dot{M}_{\mathrm{out}}$ and $\Gamma$
is found for the models with $\Gamma \lesssim 2$. The new set of models also show a
saturation of $\dot{M}_{\mathrm{out}}$ beyond $\Gamma \gtrsim 2$ as in
the $T_{\mathrm{o}}=2\times 10^6$~K (fixed) models, and exhibit the
disc wind like solutions (cf.~Section~\ref{sub:Disk-Wind-Like}).  We
applied a simple power-law fit for the points with $\Gamma \lesssim 2$
(Models~28, 29, 30 and 31 in Table~\ref{tab:Model-Summary2}), and
found its index as $q=2.2\pm 0.4$ which is very similar to that of the
$T_{\mathrm{o}}=2\times 10^6$~K models and that of the CAK stellar
wind model (cf.~Table~\ref{tab:Fit-Summary}).  No outflow was found
for the models with $\Gamma \lesssim 0.3$.

The self-consistently determined
temperatures at $r=r_{\mathrm{o}}$ in the new models
depend on the polar angle. In general, the temperature decreases from
$\sim 10^7$~K in the polar direction ($\theta=0$) to $\sim 10^4$~K as
$\theta$  increases toward the equatorial plane.  At the higher polar
angles ($\theta \gtrsim 50^{\circ}$), relatively high density
accretion streams are present. The high density columns of accreting
gas have high optical depths; hence, they are very efficient for
shielding the strong ionization radiation from the central
source. This causes the lower temperatures at $r=r_{\mathrm{o}}$ for the higher $\theta$ values.

In summary, the inflow and outflow properties for the new models are
very similar to those of the low outer boundary temperature models
($T_{\mathrm{o}}=2\times 10^6$~K). The disc wind like solutions also
appear for the models with high accretion rates ($\Gamma \gtrsim$ 2).


\begin{table*}

\caption{Summary of models without $T_{\mathrm{o}}$ constrained}

{\small \label{tab:Model-Summary2}}{\small \par}

\begin{tabular}{rccccccc}
\hline 
 & $\rho_{\mathrm{o}}$ & $\dot{M}_{\mathrm{in}}\left(r_{\mathrm{i}}\right)$ & $\Gamma$ & $\dot{M}_{\mathrm{out}}\left(r_{\mathrm{o}}\right)$ & $T_{\mathrm{ave}}^{\dagger}$ & Variability & Outflow Morphology\tabularnewline
Model & $\left(10^{-21}\,\mathrm{g\, cm^{-3}}\right)$ & $\left(10^{25}\mathrm{\, g\, s^{-1}}\right)$ & $\cdots$ & $\left(10^{25}\,\mathrm{g\, s^{-1}}\right)$ & $\left(10^{6}\,\mathrm{K}\right)$ & $\cdots$ & $\cdots$\tabularnewline
\hline 
28 & 10 & 8.9$\left(0.68\right)^{\ddagger}$ & 0.52$\left(0.04\right)^{\ddagger}$ & 1.0$\left(0.032\right)^{\ddagger}$ & 1.4 & semi-steady & wide polar\tabularnewline
29 & 20 & 12$\left(0.58\right)$ & 0.71$\left(0.034\right)$ & 3.5$\left(0.52\right)$ & 1.4 & semi-steady & wide polar\tabularnewline
30 & 40 & 18$\left(1.1\right)$ & 1.1$\left(0.1\right)$ & 7.1$\left(1.4\right)$ & 3.6 & semi-steady & wide polar\tabularnewline
31 & 80 & 25$\left(2.6\right)$ & 1.4$\left(0.2\right)$ & 10$\left(2.3\right)$ & 8.0 & semi-steady & wide polar\tabularnewline
32 & 160 & 36$\left(4.9\right)$ & 2.1$\left(0.3\right)$ & 9.9$\left(2.6\right)$ & 9.8 & semi-steady & disc-wind like\tabularnewline
33 & 320 & 52$\left(2.6\right)$ & 3.1$\left(0.2\right)$ & 11$\left(0.48\right)$ & 8.5 & semi-steady & disc-wind like\tabularnewline
34 & 640 & 72$\left(0.56\right)$ & 4.3$\left(0.03\right)$ & 9.5$\left(0.19\right)$ & 13 & steady & disc-wind like\tabularnewline
\hline
\end{tabular}

\raggedright

($\dagger$)~Self-consistently determined average temperatures at
the outer boundary

($\ddagger$)~Values in brackets are the standard deviations of the
time series values. 

\end{table*}


\subsection{Assumption of flows with optically-thin UV radiation}
\label{sub:thin_UV}
\textcolor{black}{
  As briefly mentioned in Section~\ref{sub:Overview}, our model takes
  into account for the attenuation of X-ray radiation by computing
  X-ray optical depth when computing the radiative cooling/heating
  rate and radiation forces; however, we assume that the flow is
  optically thin in UV.  The attenuation of the UV radiation is not
  considered here because our model uses the
  method described by \citet{Stevens:1990} who parametrised the line
  force by the photoionization parameter $\xi$. This simplification
  speeds up our calculation tremendously since $\xi$ can be
  estimated rather 
  quickly without computing the detail ionization levels of the
  gas. The formulation of \citet{Stevens:1990} assumes that the UV
  radiation is optically thin; therefore, to be consistent with their
  original model, our model must make the same assumption. 
}

\textcolor{black}{
  We have checked the radial UV optical depths (between the inner and
  outer boundaries) of a several representative models (especially for
  those with high mass accretion rates, $\Gamma > 1$) at a few time
  slices.  For each case, the optical depths ($\tau$) have been computed
  along each polar angle ($\theta$).  We found that $\tau$ in general
  increases as $\theta$ increase. The optical depth in polar directions
  ($\theta \simless 60^{\circ}$) are almost always less than 0.1. 
  Even at larger polar angles ($60^{\circ} \simless  \theta \simless
  85^{\circ}$), $\tau < 1$. Only along a couple of  $\theta$ angles
  near the equator ($\sim 90^{\circ}$) where the relatively high
  density gas is prominent (c.f.~Fig.~\ref{fig:disk-wind-examples}),
  $\tau$ becomes greater than 1.
  From this analysis, we conclude that our assumption of optically
  thin UV radiation in our model is reasonable. 
  Since the input radiation field in our model becomes smaller as
  $\theta$ increases (c.f.~equation~\ref{eq:rad-force-final}) and it
  becomes very small along the equator, the effect of the small optical
  depths at larger $\theta$ angles found here is minimal. On
  the other hand, if the angular distribution of the radiation field
  peaks in the equatorial direction, then the UV optical depths found
  here would make some effect on our results; however, this is not the
  case considered in this work.
}

\textcolor{black}{
  If, however, the optical depth becomes very
  large ($\tau\gg1$) in the polar region in which our input radiation
  field is strongest, we would need to consider the effect of
  attenuation and scattering of the photons
  (see e.g.~\citealt*{Ostriker:1991}; \citealt{Murray:1994}; \citealt{Ohsuga:2005}). In particular,
  the scattered photons from the polar region to equatorial region
  would redistribute the angular distribution of our input radiation
  field. Furthermore, the scattered photons would heat up the
  relatively low temperature accreting gas near the equator
  (c.f.~Fig.~\ref{fig:disk-wind-examples}). This would 
  especially change the scale height of the `disc-like' accretion
  structure, and the corresponding mass-accretion rate.  To
  include the effect of multiple scattering of photons, one would need
  a proper treatment of radiative transfer. Implementing such an effect
  in our model would require a major change in our computational code,
  and this is beyond the scope of this paper.
}

\section{Conclusions}

\label{sec:Conclusions}

We have performed a set of 2-D (axisymmetric) hydrodynamical simulations
(Table~\ref{tab:Model-Summary}) to study radiation-driven outflows
from a central part ($\lesssim10$~pc) of AGN. We have extended the
radiation-driven AGN outflow model of \citet{Proga:2007b} by relaxing
the assumption of a constant accretion luminosity. 
This allows us to determine the accretion luminosity consistently with the mass accretion 
rate at the inner boundary, and consequently the two quantities are
coupled through the radiation field.  Therefore, we made another step toward
a more comprehensive self-consistent hydrodynamical 
models with radiative feedback. 
\textcolor{black}{
  Although the accretion luminosity is self-consistently determined
  in our model, the angular distribution of the radiation field is
  not self-consistently determined, but it is fixed while
  simulations proceed. The radiation always peaks in the polar
  direction and decreases significantly in the equatorial
  direction.
}

We have examined the dependency of mass outflow rates on the accretion
luminosity (Section~\ref{sub:Mdot-L-relation}) for three different
outer boundary temperatures ($T_{\mathrm{o}}=2\times10^{6}$, $2\times10^{7}$
and $2\times10^{8}$~K) by varying the outer boundary density ($\rho_{\mathrm{o}}$).
In the following, we summarise our main findings. 

We found a relatively strong correlation between the mass outflow rates 
$\dot{M}_{\mathrm{out}}$
at the outer boundary ($r=r_{\mathrm{o}}$) and the Eddington ratio $\Gamma$
(Fig.~\ref{fig:MdotOut_Gamma}). The outflows found in our models
can be divided into three groups: 1.~thermally-driven winds, 2.~radiation-driven
winds with a wide inflow, and 3.~radiation-driven wind with a
narrow equatorial inflow (disc-wind like solutions). Out of 22 models considered
here, the majority of them (19 models) produces radiation-driven
outflow/winds (Groups~2 and 3). Only 3 models produce thermally driven
winds (Grope~2: Models~15, 16 and 17). The thermally driven winds are
found only in the highest temperature models
($T_{\mathrm{o}}=2\times10^{8}$~K). The disc-wind like outflows 
are found in 4 cases (Group~3: Models~6, 7, 8 and 9) which all have
the lowest temperature at $r=r_{\mathrm{o}}$ 
($T_{\mathrm{o}}=2\times10^{6}$~K). No outflow is found for the models
with $\Gamma\lesssim0.3$ when the ambient temperature is
$T_{\mathrm{o}}=2\times10^{6}$~K, and for the models with
$\Gamma\lesssim0.2$ when $T_{\mathrm{o}}=2\times10^{7}$~K.  

For the models with radiation-driven outflows, we found that the correlation
between $\dot{M}_{\mathrm{out}}$ and $\Gamma$ can be described as
a power law. The slope of the index of the power law ($q$) is relatively
insensitive to the outer boundary temperature $T_{\mathrm{o}}$(Table~\ref{tab:Fit-Summary}).
Using the models from all three values of $T_{\mathrm{o}}$ (see above),
the index of the power law is found as $q=2.0\left(\pm0.1\right)$.
This is very similar to those of the radiation-driven stellar wind
(e.g.~\citealt{Castor:1975}) and disc wind (e.g.~\citealt{proga:1998};
\citealt{Proga:1999}) which predict $q\geq 1.67$. This confirms that
a very similar power-law scaling of mass outflow rates with the luminosity
of the continuum source works at extremely different mass regimes,
namely $M_{*}\sim1\,\mathrm{M_{\odot}}$ (cataclysmic variables),
$\sim10\,\mathrm{M_{\odot}}$ (massive stars) and $\sim10^{8}\,\mathrm{M_{\odot}}$
(AGN), which also have different types of spectral energy
distributions and the geometry of the radiation field.

As in Papers~I, II, and III, we find that the outflow is driven
from an inflow. However, here we found cases of the inflow-outflow
solution of an extreme form (Section~\ref{sub:Disk-Wind-Like})
for the lowest outer boundary temperature
models ($T_{\mathrm{o}}=2\times10^{6}$~K) but with relatively high
outer boundary densities ($\rho_{\mathrm{o}}\gtrsim2\times10^{-20}\,\mathrm{g\, cm^{-3}}$).
In these models, the inflow occur in a narrow zone near the
equatorial plane which resembles a thin accretion disc, and the outflow
occurs in a wide polar angles ($0<\theta\lesssim85^{\circ}$). This
type of the flow is very similar to a wind driven by radiation
from a luminous accretion disc
(e.g.~\citealt{proga:1998}; \citealt{Proga:1999}; \citealt{Proga:2002b}). 
In our models, the inflowing gas is very 
slowly rotating, so that  the rotation velocity of 
the inflowing disc-like structure is much smaller than the Keplerian
velocity. The outflows arise from the disc-like structure near the
equatorial plane in a wide range of the radii 
($1.4\times10^{-2}\,\mathrm{pc}\lesssim r\lesssim7\,\mathrm{pc}$).

For the models with a relatively high density set at $r=r_{\mathrm{o}}$,
a system reaches a steady state at $\Gamma>1$ (see Table~\ref{tab:Model-Summary}
and Fig.~\ref{fig:MdotOut_Gamma}). In other words, we find the models
can still accrete matter even with a super-Eddington luminosity (up
to $\Gamma\sim4$). This is a multi-dimensional effect. In our axi-symmetric
model, the radiation field or the strength of the radiation force
is not spherically symmetric but has the $\cos\theta$ dependency (equation~\ref{eq:g_rad_continnum}).
This allows gas to accrete near the equatorial plane even with a super-Eddington
luminosity ($\Gamma>1$), as demonstrated in Fig.~\ref{fig:g_rad_vs_g_gravity}.
Based on this figure, the correlation between $\dot{M}_{\mathrm{out}}$
and $\Gamma$ (for the radiation driven outflow cases) is expected
to continue up to $\Gamma \sim 10$; however, we are not be
able to confirm this due to numerical difficulties in handling very
strong shocks that occur in the flows in the models with $\Gamma>5$ 
(for $T_{\mathrm{o}}=2\times10^{7}$ and $2\times10^{8}$~K models).  In the range
of $\Gamma$ values we have explored, we do not observe a complete
shutdown of the accretion flow even at a super-Eddington luminosity,
which is expected in a spherically symmetric model.
The accretion is still self-regulating, but not as a spherical model
predicts.  

When we relax the assumption of a constant temperature at the outer
boundary and self-consistently determine the temperatures there, the
$\dot{M}_{\mathrm{out}}$--$\Gamma$ relation 
becomes very similar to that of the models with the lowest outer
boundary temperature, $T_{\mathrm{o}}=2\times10^{6}$~K
(Fig.~\ref{fig:To_test}).   The high accretion rate models ($\Gamma
\gtrsim 2$) show disc wind like solutions, as seen in the high
accretion rate models with $T_{\mathrm{o}}=2\times10^{6}$~K
(Section~\ref{sub:Disk-Wind-Like}).

The following is a list of some steps for our model that shall
be considered in the future: (i)~include the effect of scattered
photons which might enhance heating of the cold equatorial inflow, 
(ii)~adjust the continuum spectral energy distribution based on the
mass and the luminosity of the system, and (iii)~extend the
computational domain up to a few 100~pc scale to capture the gas
dynamics in the narrow line regions of AGN, and
(iv)~include dust.

In addition, we shall incorporate the results
from smaller scale simulations (e.g.~disc-wind models by \citealt{Proga:1999})
since not all the gas which accretes though our inner boundary (which
is relatively large i.e.~$r=0.014$~pc) would reach the central
SMBH due to the wind from the accretion disc. 

\section*{Acknowledgements} 

Authors thank the anonymous referee for constructive
comments and suggestions for improving the clarity of the manuscript.
We thank Prof.~J.~Ostriker for very helpful discussion and comments. 
This work was supported by NASA through grant HST-AR-11276 from the
Space Telescope Science Institute, which is operated by the Association
of Universities for Research in Astronomy, Inc., under NASA contract
NAS5-26555. A significant fraction of our simulations were performed
on a SUN computer system funded by President of UNLV, D. B. Ashley
through an Infrastructure Award to the Astronomy Group at UNLV. This
work was also supported by the National Center for Supercomputing
Applications under AST070036N which granted the accesses to the Intel
64 Linux Cluster Abe. Authors are grateful for the original developers
of {\sc ZEUS-MP} for making the code publicly available. 


\end{document}